\documentclass{article}
\usepackage{iclr2027_conference,times}

\usepackage{amsmath,amsfonts,bm}

\def\eqref#1{equation~\ref{#1}}

\def\1{\bm{1}}

\DeclareMathAlphabet{\mathsfit}{\encodingdefault}{\sfdefault}{m}{sl}
\SetMathAlphabet{\mathsfit}{bold}{\encodingdefault}{\sfdefault}{bx}{n}

\usepackage{hyperref}
\usepackage{url}
\usepackage{multicol}
\usepackage{multirow}
\usepackage{booktabs}
\usepackage{wrapfig}
\usepackage{bbold}
\usepackage{graphicx}
\usepackage{caption}
\usepackage{xcolor}
\usepackage[most]{tcolorbox}
\usepackage{colortbl}
\usepackage{enumitem}
\usepackage{algorithm}
\usepackage{algpseudocode}
\usepackage{float}
\usepackage{etoc}

\usepackage[table,dvipsnames]{xcolor}
\definecolor{cwA}{HTML}{7CA9B6}
\definecolor{cwB}{HTML}{A0D2D6}
\definecolor{cwC}{HTML}{C8EDEF}
\definecolor{cwD}{HTML}{E1F7F8}

\title{\includegraphics[height=1em]{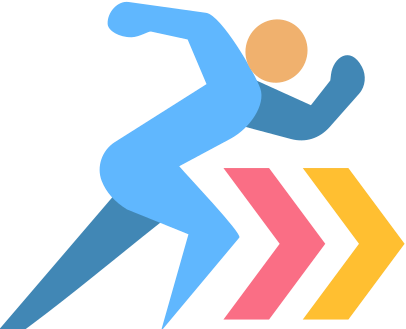} SPRINT: Single-Step Generative Recommendation via Average Probability Velocity}

\author{Zhuo Cai$^{1}$ \quad
Shoujin Wang$^{1}$\thanks{Corresponding author} \quad
Peilin Zhou$^{2}$ \quad
Min Xu$^{1}$ \quad
Julian McAuley$^{3}$ \quad
Fang Chen$^{1}$ \quad \\
$^{1}$University of Technology Sydney \quad
$^{2}$New York University Abu Dhabi\quad \\
$^{3}$University of California, San Diego \quad
}

\iclrfinalcopy
\begin{document}
\lhead{Preprint.}

\maketitle

\begin{abstract}
Semantic ID (SID) based generative recommendation represents each item as a sequence of discrete tokens, and recommends by generating the SID of the item a user would like to interact with. Both dominant paradigms in this domain generally pay for generation token by token: \textit{autoregressive models} decode the tokens left-to-right, while \textit{non-autoregressive models} decode in parallel yet still need multiple rounds of refinement to stay competitive. Therefore, both generally spend multiple forward passes per item, a cost that is prohibitive in latency-sensitive recommender systems. We ask whether an item can be generated in a single forward pass, and answer it through a new perspective which we call \emph{average probability velocity}. We view SID generation as a flow of token generation probabilities and characterize it by its average velocity over the whole generation process. We prove that this average velocity is fully determined by the average generation probability of each token. Therefore, we directly parameterize and learn the probabilities of all tokens in a single forward pass with a bidirectional Transformer. As these probabilities are generated independently across positions and the coherence among tokens is lost, we further design a dual-level flow contrastive objective to restore the coherence among an item's tokens. It contrasts the target SID against negative SIDs at both the token and SID levels. The token level ranks the generation probabilities of the target tokens above those of negative SIDs, while the SID level scores the tokens of each SID as a whole item for capturing token coherence of each item. Extensive experiments on 8 real-world datasets show that our model not only generates recommendations far more efficiently ($8.39-10.04\times$ speedup over the second-fastest AR/NAR method) but also attains superior recommendation accuracy ($7.77\%$ average improvement over the second-best). Our code is available at \url{https://github.com/iamZhuoCai/SPRINT}.
\end{abstract}

\etocdepthtag.toc{mtchapter}

\section{Introduction}

Generative recommendation (GR) with semantic IDs (SIDs) tokenizes each item into a short sequence of discrete tokens (\textit{i.e.}, an SID), such as $[\texttt{a\_2},\texttt{b\_24},\texttt{c\_182},\texttt{d\_78}]$, and recommends by generating the SID of the next item a user is interested in~\citep{rajput2023recommender, wang2024learnable}. Since an SID spans $L$ tokens (typically $3$--$4$), the cost of a single recommendation is determined by the number of forward passes required to produce these $L$ tokens. Autoregressive (AR) GR decodes them left-to-right, one token per pass, and therefore requires $L$ passes~\citep{rajput2023recommender,Hou2025Action}. Non-autoregressive (NAR) GR, most prominently masked diffusion-based~\citep{shi2025llada,liu2026diffgrm,mu2026masked,shah2025masked}, removes the left-to-right constraint and can, in principle, generate all $L$ tokens in a single pass. However, in pactice, these methods generally still need to generate the tokens one by one to build the conditioning among tokens~\citep{shi2025llada}.

\begin{figure*}[t]
\centering
\begin{minipage}[c]{0.42\textwidth}
  \centering
  \includegraphics[width=\linewidth]{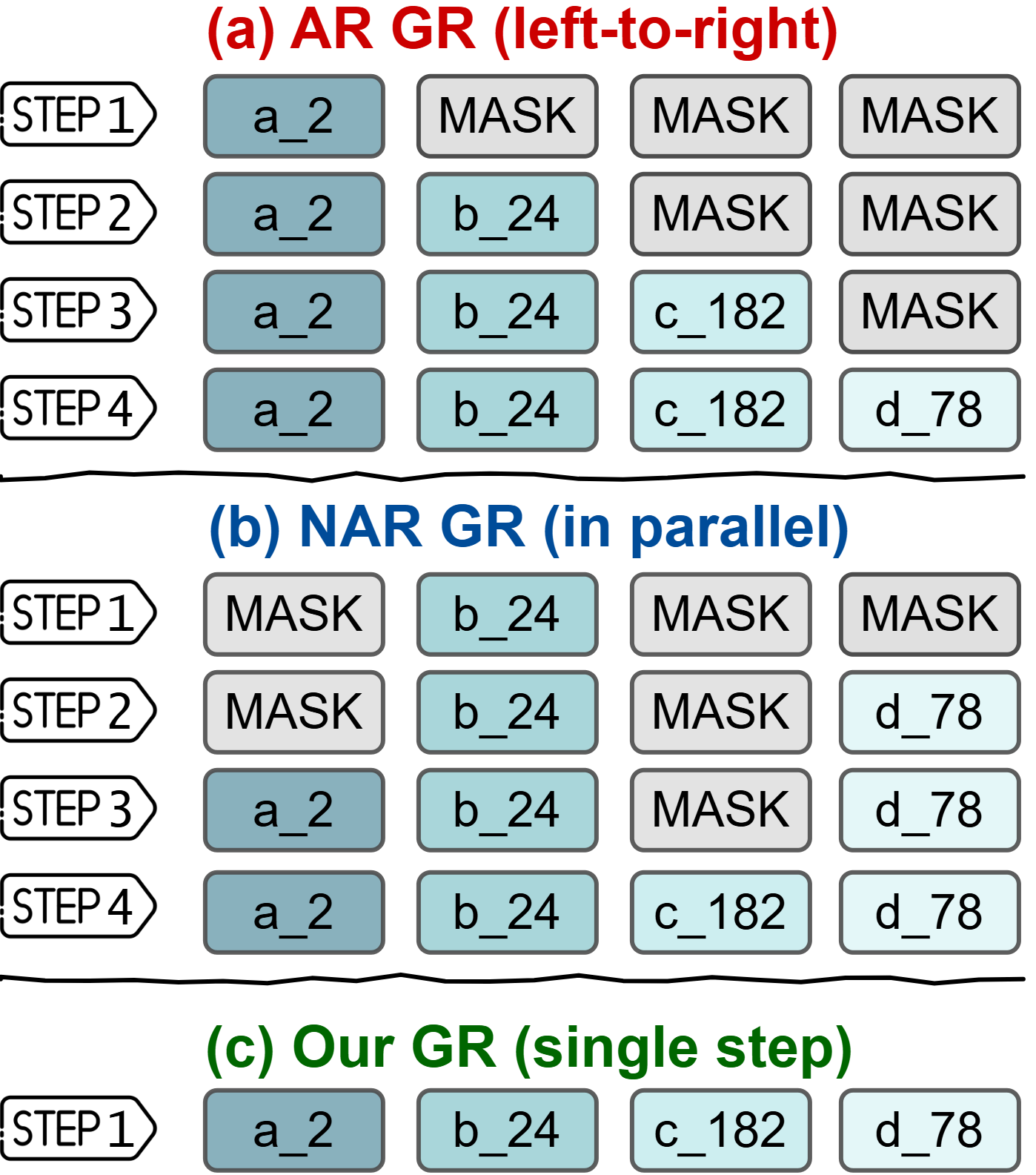}
  \vspace{-1em}
  \captionof{figure}{Comparison of AR, NAR, and our generative
    recommendation (GR) models. [\colorbox{cwA}{a\_2},\colorbox{cwB}{b\_{24}},\colorbox{cwC}{c\_{182}},\colorbox{cwD}{d\_{78}}] is the token sequence (\textit{i.e.}, SID) of an item.}
  \label{fig:demo}
\end{minipage}
\hfill
\begin{minipage}[c]{0.56\textwidth}
  \centering
  \footnotesize
  \setlength{\tabcolsep}{2.5pt}
  \renewcommand{\arraystretch}{1.3}
  \captionof{table}{Single-step vs.\ four-step generation performance of
    existing masked diffusion-based NAR GR methods. LLaDA-Rec \citep{shi2025llada} and DiffGRM \citep{liu2026diffgrm} are two representative and state-of-the-art masked diffusion-based methods. In these methods, each item is indexed by 4 discrete tokens. ``R'' and ``N'' are the abbreviations of Recall and NDCG.}
  \label{tab:pilot_experiments}
  \begin{tabular}{llcccc}
  \toprule
  \multirow{2}{*}{Methods} & \multirow{2}{*}{Generation}
    & \multicolumn{2}{c}{Scientific}
    & \multicolumn{2}{c}{Instruments}\\
  \cmidrule(lr){3-4} \cmidrule(lr){5-6}
   & steps & R@10 & N@10 & R@10 & N@10 \\
  \midrule
  \multirow{3}{*}{LLaDA-Rec}
    & 4-step    & 0.0474 & 0.0256 & 0.0623 & 0.0337 \\
    & \cellcolor{gray!30}1-step    & \cellcolor{gray!30}0.0033 & \cellcolor{gray!30}0.0019 & \cellcolor{gray!30}0.0023 & \cellcolor{gray!30}0.0013 \\
    & Drop (\%) & 93.04\% & 92.58\% & 96.31\% & 96.14\% \\
  \midrule
  \multirow{3}{*}{DiffGRM}
    & 4-step    & 0.0605 & 0.0337 & 0.0659 & 0.0356 \\
    & \cellcolor{gray!30}1-step    & \cellcolor{gray!30}0.0066 & \cellcolor{gray!30}0.0042 & \cellcolor{gray!30}0.0152 & \cellcolor{gray!30}0.0069 \\
    & Drop (\%) & 89.09\% & 87.54\% & 76.93\% & 80.62\% \\
  \bottomrule
  \end{tabular}
\end{minipage}
\end{figure*}

Either way, generating a single item costs $L$ forward passes, further multiplied by beam width for a ranked list. This forms an efficiency bottleneck in latency-sensitive recommender systems, which must serve requests within a tight latency budget: added delay degrades user engagement and inflates serving cost at scale \citep{guo2026mlps}. A natural question is whether we can pay just one forward pass to derive an SID consisting of $L$ tokens. Simply running an NAR model for a single step does not work. As Table~\ref{tab:pilot_experiments} shows, collapsing state-of-the-art masked-diffusion GR to one step erases up to $96.31\%$ of its accuracy (consistent with the results in \citep{shi2025llada}), because tokens decoded independently lose the conditioning that keeps an SID coherent. Existing single-step designs avoid this collapse only at a price. RPG \citep{hou2025generating} predicts all positions at once using a multi-token prediction objective, but it must inflate the SID to $16$--$64$ tokens for expressiveness, enlarging the embedding table to $4-16\times$ that of the $4$-token SIDs used by most NAR models \citep{shi2025llada, liu2026diffgrm}. OneGR \citep{wang2026one} produces position-wise scores in one pass, but it then runs an A$^\ast$ search \citep{hart1968formal} whose look-ahead expansion is itself time-consuming.

A potential remedy is flow matching \citep{Lipman2023flow}, which reduces the many sampling steps of diffusion while preserving generation quality, and whose recent one-step variants \citep{geng2025mean, geng2026improved} attain single-step generation by modeling a flow's average velocity over a finite interval. However, directly applying it to build a single-step GR is nontrivial due to two challenges (CHs). \textbf{CH1: How to attain single-step generation over the discrete token space of SIDs?} Existing single-step flow matching models \citep{geng2025mean, geng2026improved} operate in continuous space and recover the average velocity through an identity that differentiates the velocity with respect to a continuous state, which is undefined over the categorical space of SIDs. Meanwhile, existing discrete flow matching \citep{gat2024discrete, Campbell2024generative} generally requires multi-step iterative refinement. \textbf{CH2: How to maintain the inter-token coherence of an SID under the single-step paradigm?} When all $L$ tokens are decoded independently in one step, the inter-token coherence that separates the matched SID from other SIDs and arbitrary token combinations is discarded, which lowers per-position accuracy and yields inaccurate recommendation generation.

To address these challenges, in this paper, we present an effective and efficient \textbf{S}ingle-ste\textbf{P} gene\textbf{R}at\textbf{I}ve recomme\textbf{N}da\textbf{T}ion model, \textbf{SPRINT}. \textbf{To address CH1}, we build SPRINT from a new perspective called \textit{average probability velocity}. We view SID generation as a flow of token generation probabilities. At each position, the generation probability flow moves from the mask token to the target token. The rate of this movement is probability velocity. Multi-step models follow its instantaneous value and thus generate the SID token by token. We instead characterize the flow by its average velocity over the whole generation process. We theoretically prove that this average velocity is fully determined by the average generation probability of each token at each position. SPRINT therefore directly parameterizes and learns this probability with a bidirectional Transformer, which generates the probabilities of all tokens in an SID from the user history in a single forward pass. \textbf{To address CH2}, we design a dual-level flow contrastive objective, which contrasts the target SID against the SIDs of non-interacted items at two levels. At the token level, it raises the generation probabilities of the target tokens over those of negative SIDs, sharpening the prediction at each position. At the SID level, it scores each SID as a whole item rather than as separate tokens, and ranks the target SID above negative ones. This teaches the model which token combinations form the target item, restoring the inter-token coherence within a single step.

Beyond being single-step, SPRINT is concise by design, which brings two further advantages in efficiency. First, it keeps a \textbf{compact $4$-token SID}, in contrast to the $16$--$64$ tokens required by some other single-step methods \citep{hou2025generating,xia2026unleash}, which enlarge the embedding table by $4$--$16\times$. Second, its \textbf{architecture is lightweight} (a bidirectional Transformer with a $1$-layer encoder and a $4$-layer decoder) with no auxiliary mask-scheduling (\textit{e.g.}, history-aware mask position allocation \citep{mu2026masked}) or data-augmentation machinery (\textit{e.g.}, on-policy coherent noising \citep{liu2026diffgrm}). Extensive experiments on 8 real-world datasets show that SPRINT decodes $8.39$--$10.04\times$ faster than the second-fastest AR/NAR method while improving accuracy by $7.77$\% on average across all datasets and metrics over the second-best. \textbf{Our main contributions are summarized as:} \textbf{(1)} We are the first to model SID generation from the perspective of average probability velocity, casting the generation of an item's SID as the average velocity of the probability flow over token generation, generating SID in a single step; \textbf{(2)} We propose a novel dual-level flow contrastive objective to restore inter-token coherence within a single step; \textbf{(3)} We provide theoretical analysis and extensive experiments on 8 datasets, which verify both the efficiency and the accuracy of SPRINT.

\vspace{-0.5em}
\section{Related Work}
\vspace{-1em}
We briefly review related work here; please refer to Appendix \ref{sec:more_related_work} for more related work.

\textbf{Autoregressive GR.} Autoregressive (AR) GR decodes the tokens of an SID from left to right, imposing a strict sequential
dependency. TIGER~\citep{rajput2023recommender} pioneered this paradigm by quantizing item embeddings into semantic IDs (SIDs) via residual quantization~\citep{lee2022autoregressive} and decoding them
token-by-token. Subsequent work extends it through semantic-collaborative alignment~\citep{zheng2024adapting, deng2025onerec}, industrial-scale transducers~\citep{Zhai2024Actions}, and contextual tokenization~\citep{Hou2025Action}. Whatever their differences, all of these methods inherit the cost of left-to-right decoding: an SID of $L$ tokens requires $L$ forward passes.

\textbf{Non-autoregressive GR.} Non-autoregressive (NAR) GR generates the tokens of an item's SID in parallel, relaxing the left-to-right sequential dependency of AR decoding~\citep{ren2024non}. Masked diffusion-based GR \citep{shi2025llada, liu2026diffgrm, mu2026masked, shah2025masked} is the most widely adopted NAR paradigm. Although NAR GR can decode from arbitrary positions, it still requires token-by-token refinement to capture token dependency to competitive performance~\citep{shi2025llada}, thus still require $L$ forward passes. RPG~\citep{hou2025generating} and OneGR~\citep{wang2026one} instead predict all tokens simultaneously, but they either greatly enlarge the embedding table or rely on an A$^*$ search~\citep{hart1968formal} to select tokens sequentially at the
generation stage. In contrast, we propose a theoretically principled, effective, and efficient single-step GR model.

\textbf{Flow matching.}
Flow matching~\citep{Lipman2023flow, Liu2023flow} learns a time-dependent velocity field transporting a prior to the data distribution. Although it reduces the number of steps required by diffusion, it models the instantaneous velocity and therefore still needs to solve the generation trajectory over multiple steps. To remove this cost, single-step flow matching models~\citep{geng2025mean, geng2026improved, kornilov2024optimal} instead learn the average velocity over a finite interval. However, they are all defined only in continuous space. Discrete flow matching~\citep{Campbell2024generative, gat2024discrete} does extend the framework to discrete data, but remains built on instantaneous transition rates and thus still entails iterative sampling. Flow matching has also been introduced to recommender systems~\citep{Shi2026fave, ye2026gaussian}, yet none of them operates in the discrete SID-based generative recommendation domain. In this paper, we bridge this gap by developing a single-step GR model from the perspective of average probability velocity.

\section{SPRINT: Single-step Generative Recommendation}
\subsection{Problem Formulation and Overview}
\textbf{Sequential recommendation.} Existing generative recommendation (GR) generally follows the sequential recommendation paradigm \citep{rajput2023recommender}. Specifically, given a user's chronologically ordered interaction history $\mathcal{H}=[i_1,i_2,\cdots, i_n]$, the goal is to predict the next item $i_{n+1}$ the user would like to interact with, ultimately producing a ranked list of $K$ candidates.

\textbf{Items as discrete token sequences (SID).} Following the GR paradigm, each item is not an atomic index but a short tuple of $L$ discrete tokens $[c^1_i,c^2_i,\cdots,c^L_i]$, \textit{i.e.}, a semantic ID (SID). A user's
interaction history is then represented as $\mathcal{H}=[c^1_1,\cdots,c^L_1,\cdots,c^1_n,\cdots,c^L_n]$. This discretization can be realized by various vector quantization techniques, such as RQ-VAE \citep{lee2022autoregressive,rajput2023recommender} and optimized product quantization (OPQ) \citep{ge2013optimized,jegou2010product}. We adopt OPQ following previous work \citep{hou2025generating, liu2026diffgrm, wang2026one}, because
single-step generation does not decode a token conditioned on preceding ones, which is precisely what residual tokenizers assume. We empirically verify this choice in Section~\ref{sec:ablation}, where replacing OPQ with RQ-VAE or RQ-Kmeans consistently degrades accuracy.

\textbf{Generative recommendation as next SID generation.} Recommending an item then means generating its SID $[c^1_{n+1},c^2_{n+1},\cdots,c^L_{n+1}]$ conditioned on $\mathcal{H}$. Dominant paradigms decode this sequence either autoregressively, one token at a time from left to right and thus in $L$ forward passes \citep{rajput2023recommender}, or, more recently, in parallel or in arbitrary order via masked diffusion \citep{liu2026diffgrm,shi2025llada,mu2026masked}. However, as shown in Table~\ref{tab:pilot_experiments}, the latter still requires $L$ forward passes to attain competitive recommendation accuracy.

\textbf{Our goal: single-step generation.} We aim to produce the full target SID in a single forward pass. To this end, we cast decoding as a discrete flow from an all-mask source to the target SID, and rather than integrating that flow with many small steps, we learn an \emph{average
probability velocity} that jumps from source to target in one shot. The remainder of this section develops (i) Setup and notations (\S\ref{sec:setup}); (ii) Average probability velocity for single-step SID generation (\S\ref{sec:apv}); (iii) Dual-level flow contrastive training and inference (\S\ref{sec:dual}).

\subsection{Setup and Notations}
\label{sec:setup}

\textbf{Intuition.} Generating a recommendation starts from an all-mask SID $[\mathbb{m},\cdots,\mathbb{m}]$ in which no position has been decided, and ends at the SID of the item a user would like to interact with. This can be viewed as a process indexed by a time $t\in[0,1]$ along which positions gradually commit to their target tokens. Existing GR realizes this process step by step, whereas we ask how far the whole process moves on average over the interval, so that it can be completed in one jump.

\textbf{SID generation probability path.} We treat SID generation as transporting probability mass from a source distribution $p$ to a target distribution $q$ over discrete token sequences of length $L$. Let $p_0(x)=p$ and $p_1(x)=q$, and let a time-indexed path $p_t(x)$, $t\in(0,1)$, interpolate between them. In our setting the source is the fully masked sequence $[\mathbb{m},\cdots,\mathbb{m}]$ of length $L$ and the target is the SID $[c^1_{n+1},\dots,c^L_{n+1}]$ of the target item $i_{n+1}$. Given a coupling $\pi(x_0,x_1)$ of source and target endpoints, which in our case couples the all-mask start $x_0=[\mathbb{m},\cdots,\mathbb{m}]$ with the ground-truth next item's SID $x_1=[c^1_{n+1},\dots,c^L_{n+1}]$, the marginal path factorizes over positions through a conditional path:
\begin{equation}
p_t(x)=\sum_{x_0,x_1}p_t(x|x_0,x_1)\pi(x_0,x_1),\qquad
p_t(x|x_0,x_1)=\prod_{l=1}^{L}p_t(x^l|x_0,x_1).
\end{equation}
The endpoints are pinned by delta functions \citep{gat2024discrete}, $p_0(x|x_0,x_1){=}\delta_{x_0}(x)$ and  $p_1(x|x_0,x_1){=}\delta_{x_1}(x)$, where $\delta_{y}(x)=\prod_l\delta_{y^l}(x^l)$ and $\delta_{y^l}(x^l)=\mathbb{1}[x^l=y^l]$.

\textbf{Convex-interpolant path.} A simple and effective per-token path linearly
mixes the two endpoints under a scalar scheduler $\kappa_t$ (\textit{e.g.},
$\kappa_t=t$):
\begin{equation}
    p_t(x^l|x_0,x_1)=(1-\kappa_t)\delta_{x_0}(x^l)+\kappa_t\delta_{x_1}(x^l), \qquad \kappa_0=0,\kappa_1=1, \dot{\kappa}_t\geq0,
\end{equation}
where $\kappa_t$ measures how much of the target item's SID has
been decided: at $\kappa_t=0$ every SID position is still masked and nothing is
committed, and at $\kappa_t=1$ all positions have committed to the tokens of the
target SID. Here $\dot{\kappa}_t:=\frac{d\kappa_t}{dt}$ is the
instantaneous rate at which positions switch from mask to their target token.

\textbf{Instantaneous probability velocity.} As $t$ advances from $0$ to $1$, each SID position occasionally jumps from its current value to its target value, so that the target item is progressively revealed and running the process to $t=1$ yields the complete SID. What governs these jumps is the probability velocity field $v_t$: $v^l_t(x^l,z)$ is the rate at which position $l$, currently holding value $z^l$, jumps to value $x^l$. Concretely, over an infinitesimal step $h>0$, a token at $X^l_t=z^l$ has next-state distribution $X^l_{t+h}\sim\delta_{z^l}(\cdot)+hv^l_t(\cdot,z)$. The marginal velocity $v_t$ cannot be computed directly, so we construct it as the marginalization of a conditional velocity, for which we introduce the following interpolant:
\begin{equation}
v^l_\tau(x^l,z|x_0,x_1)=v^l_\tau(x^l,z|x_1)
=\frac{\dot{\kappa}_\tau}{1-\kappa_\tau}\big[\delta_{x_1}(x^l)-\delta_z(x^l)\big].
\label{eq:inst}
\end{equation}
The bracket $[\delta_{x_1}-\delta_z]$ adds probability mass onto the target $x_1$ and removes it from the current state $z$, that is, it pushes each position away from where it currently sits and toward the token of the item being generated, while the scalar $\dot{\kappa}_\tau/(1-\kappa_\tau)$ sets the pace toward $x_1$. Since $x_0$ is always the all-mask sequence, it can be omitted from the conditioning. The derivation of this interpolant is provided in Appendix~\ref{derivation_interpolant}, and we prove its validity as a CTMC generator \citep{gat2024discrete, Campbell2024generative} in Appendix~\ref{sec:validity}. Standard discrete flow matching would learn $v_\tau$ and integrate it with many small jump steps; we next show how to collapse it into a single step.

\begin{figure}
    \centering
    \includegraphics[width=1\linewidth]{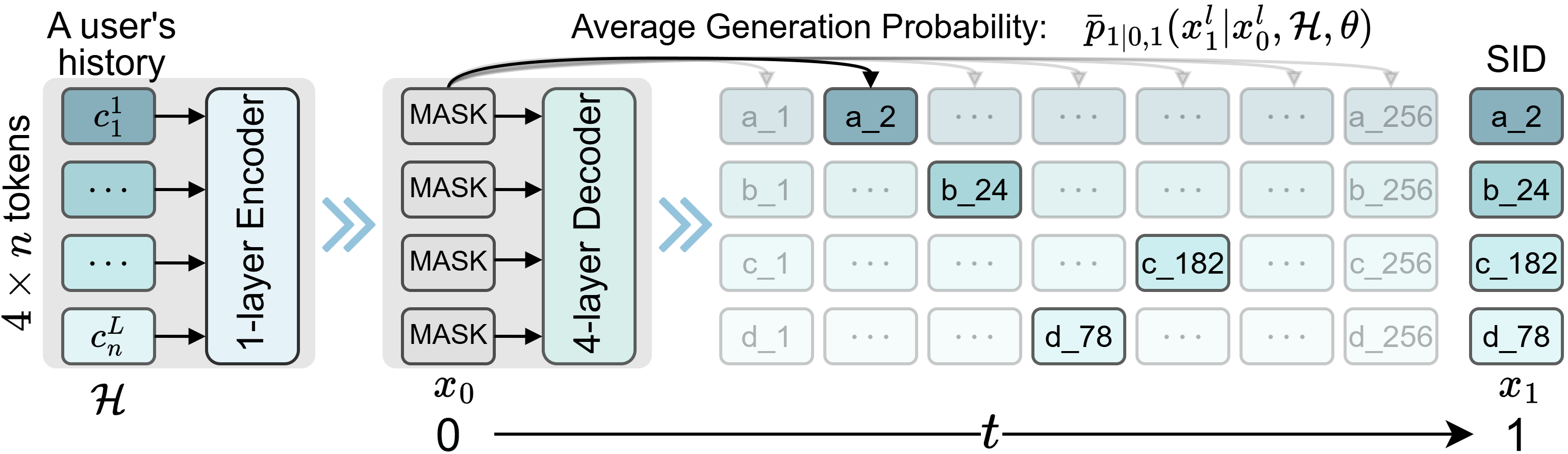}
    \caption{Illustration of SPRINT's single-step generation via average probability velocity. We here only explicitly present the generation principle of the first token of an item. In fact, all tokens are generated simultaneously under the same principle.}
    \label{fig:placeholder}
    \vspace{-1em}
\end{figure}

\subsection{Average Probability Velocity for Single-Step GR}
\label{sec:apv}
\textbf{Average probability velocity.} We learn the average probability velocity over a finite interval $[r,t]$, defined as the time-averaged instantaneous probability velocity:
\begin{equation}
\bar{v}^l_{r,t}(x^l,z)=\frac{1}{t-r}\int_r^t v^l_\tau(x^l,z)\,\mathrm{d}\tau .
\label{eq:avg}
\end{equation}
Writing this in conditional form and substituting Eq.~\ref{eq:inst}, we obtain
\begin{equation}
\label{eq:conditional_average_velocity}
\bar{v}^l_{r,t}(x^l,z|x_1)
=\frac{1}{t-r}\left[\int_r^t\frac{\dot{\kappa}_\tau}{1-\kappa_\tau}\mathrm{d}\tau\right]
\big[\delta_{x_1}(x^l)-\delta_z(x^l)\big]
=\lambda_{r,t}\big[\delta_{x_1}(x^l)-\delta_z(x^l)\big],
\end{equation}
with $\lambda_{r,t}=\frac{1}{t-r}\ln\frac{1-\kappa_r}{1-\kappa_t}$ \footnote{Note that $\lambda_{r,t}\to\infty$ as $t\to 1$, which is the expected behaviour of a CTMC rate when the transition must complete by the terminal time~\citep{Campbell2024generative}. It serves only as an intermediate quantity in the derivation and is never used during optimization or generation.} (complete derivation in Appendix~\ref{sec:con_ave_pro_vel}). This is precisely where our formulation departs from continuous mean-velocity methods: because the direction of transport $[\delta_{x_1}-\delta_z]$ does not vary with $\tau$, averaging the
velocity reduces to averaging a scalar rate, so the discrete convex interpolant makes the average probability velocity integrable in closed form and $\lambda_{r,t}$ is obtained without the Jacobian--vector-product identity used in the continuous case.

\textbf{Marginal average probability velocity.} Since the target $x_1$ is
unknown at inference, we construct the marginal counterpart:
\begin{equation}
\label{eq:marginal_average_velocity}
\bar{v}^l_{r,t}(x^l,z)
=\frac{1}{t-r}\int_r^t\frac{\dot{\kappa}_\tau}{1-\kappa_\tau}
\big[p_{1|\tau}(x^l|z)-\delta_z(x^l)\big]\mathrm{d}\tau
=\lambda_{r,t}\big[\bar{p}_{1|r,t}(x^l|z)-\delta_z(x^l)\big],
\end{equation}
where the average (denoising) posterior is
$\frac{\dot{\kappa}_\tau}{1-\kappa_\tau}$-weighted time average of instantaneous posterior:
\begin{equation}
\bar{p}_{1|r,t}(x^l|z):=
\frac{\int_r^t\frac{\dot{\kappa}_\tau}{1-\kappa_\tau}p_{1|\tau}(x^l|z)\,\mathrm{d}\tau}
{\int_r^t\frac{\dot{\kappa}_\tau}{1-\kappa_\tau}\,\mathrm{d}\tau}.
\label{eq:avgpost}
\end{equation}
This makes explicit what the network must learn: not the posterior at any single time, but the interval-averaged one. $\bar{p}_{1|r,t}(x^l|z)$ is the model's belief about which item to recommend, read out position by position given a partially unmasked SID $z$. It predicts the token that position $l$ of the target item should take. This is therefore the recommendation predictor itself, and single-step generation reads the recommendation straight out of it. Refer to Appendix \ref{sec:marginal-average-vel} for a complete derivation.

\textbf{Parameterization.} We approximate $\bar{p}_{1|r,t}(x^l|z)$ by a network
with parameters $\theta$. Since recommendation must be conditioned on the user's
historical interactions, the network takes both $\mathcal{H}$ and the starting
state (at $r$) as input, and we write $\bar{p}_{1|r,t}(x^l|z,\mathcal{H},\theta)$. The
backbone is a lightweight bidirectional Transformer: a $1$-layer encoder embeds the user history $\mathcal{H}$, and a $4$-layer decoder attends
to the encoder output from $L$ mask queries, one per SID position, and emits for
each position $l$ a distribution over the $M$ tokens of the corresponding
codebook $\mathcal{C}^l$.

\textbf{Time sampling.} To align training with single-step inference while still
learning a genuine interval-indexed average velocity, we fix the terminal time
$t=1$ for all training samples, since one-step generation always targets the
clean SID at $t=1$. We then sample the source time $r$ in a mixed manner: with
probability $\rho$ we set $r=0$, i.e., the fully masked source $[\mathbb{m},\cdots,\mathbb{m}]$ that
exactly matches the inference condition; with probability $1-\rho$ we draw
$r\sim\mathcal{U}(0,1)$, so that the source $x_r$ is a partially committed SID in
which each position independently holds its target token with probability
$\kappa_r$ and the mask token otherwise. The $\rho$-fraction anchors the precise
single-step map used at inference, while the remaining fraction makes
$\bar{p}_{1|r,1}$ a non-degenerate function of $r$ rather than a single fixed
denoiser, and additionally supplies conditional supervision in which the model
predicts the remaining positions of an item given those already committed.

Training should align the marginal average probability velocity in Eq. \ref{eq:marginal_average_velocity}  with its conditional counterpart in Eq. \ref{eq:conditional_average_velocity}, which amounts to aligning
$\lambda_{r,t}\bar{p}_{1|r,t}(x^l|z,\mathcal{H},\theta)$ with $\lambda_{r,t}\delta_{x_1}(x^l)$. To
achieve this, a natural choice is per-token cross-entropy loss:
\begin{equation}
-\sum\nolimits_{l=1}^{L}\log\bar{p}_{1|r,t}(x^l_1|x_r,\mathcal{H},\theta),
\label{eq:ce}
\end{equation}
where $z=x_r$ is the state of the generated SID at time $r$, $x_1$ is target item's SID, and $x^l_1$ is its $l$-th token. Following \citep{gat2024discrete}, we omit $\lambda_{r,t}$ during training since it is shared by two terms.

\textbf{From per-token likelihood to contrastive supervision.} Eq.~\ref{eq:ce}
draws supervision only from the tokens of target item's SID $x_1=[c^1_{n+1},\dots,c^L_{n+1}]$. In recommendation, however, non-target items are informative signals in their own right, since what a user does not interact with shapes the ranking as much as what they do. We therefore replace Eq.~\ref{eq:ce} with a flow contrastive loss:
\begin{equation}
-\log\frac{\exp\big(\sum_{l=1}^{L}\log\bar{p}_{1|r,t}(x^l_1|x_r,\mathcal{H},\theta)\big)}
{\sum_{c\in\mathcal{S}}\exp\big(\sum_{l=1}^{L}\log\bar{p}_{1|r,t}(c^l|x_r,\mathcal{H},\theta)\big)},
\label{eq:cl}
\end{equation}
where $c$ is the SID of an item and $\mathcal{S}$ is a sampled SID set (a target SID and $N$ randomly sampled SIDs of non-interacted items). 

\subsection{Dual-level Flow Contrastive Training and Inference}
\label{sec:dual}

The objective in Eq.~\ref{eq:cl} calculates the generation probability flow of SIDs as a product of per-token beliefs. Under single-step generation, all $L$ tokens are emitted simultaneously and therefore independently, so this factorized score cannot express how the tokens of an SID constrain one another: positions that are each individually plausible may combine into a tuple that indexes inaccurate items. Modeling this inter-token dependency is thus a second requirement of single-step GR, and we meet it by extending Eq.~\ref{eq:cl} into a dual-level flow contrastive loss:
\begin{equation}
\label{eq:dual-level-loss}
\mathcal{L}=-\sum_{(\mathcal{H},x_1)\in \mathcal{D}}\Biggl[\log\frac{
\exp\left(\sum_{l=1}^{L}\log\bar{p}_{1|r,t}(x_1^{l}\mid x_r,\mathcal{H},\theta)\right)
}{\sum_{c\in\mathcal{S}}\exp\left(\sum_{l=1}^{L}\log\bar{p}_{1|r,t}(c^{l}\mid x_r,\mathcal{H},\theta)\right)}+\log\frac{\exp\!\left(\langle\mathbf{u},\mathbf{v}_{x_1}\rangle\right)}{\sum_{c\in\mathcal{S}}\exp\left(\langle\mathbf{u},\mathbf{v}_{c}\rangle\right)
}\Biggr],
\end{equation}
where $\langle\cdot,\cdot\rangle$ denotes cosine similarity. $\mathbf{u}=\mathrm{MLP}\big(\mathrm{Cat}(\mathbf{h}^1,\dots,\mathbf{h}^L)\big)$ serves as the representation of the user/sequence. The $L$ decoder hidden states $\{\mathbf{h}^l\}_{l=1}^{L}$ are produced by $L$ mask queries attending to the encoded history $\mathcal{H}$, so each $\mathbf{h}^l$ already carries the model's belief, conditioned on the user's history, about what token position $l$ of the target item should take. Concatenating them and passing the result through an $\mathrm{MLP}$ therefore yields a single vector that summarizes the target item the user is predicted to interact with, rather than $L$ separate per-position beliefs. The item-side representation of an SID $c$ is $\mathbf{v}_c=\mathrm{MLP}\big(\mathrm{Cat}(\mathbf{e}_{c^1},\dots,\mathbf{e}_{c^L})\big)$, where $\mathbf{e}_{c^l}\in\mathbb{R}^{d_m}$ is the embedding of token $c^l$.

The two levels play complementary roles. The token-level term optimizes the marginal prediction at each position and is what makes individual tokens accurate. The SID-level term scores the $L$ tokens jointly as one object and pulls this joint assignment toward the item the user actually interacts with, which is exactly the dependency that independent decoding discards. Because $\mathbf{v}_c$ depends only on an item's SID and not on the user, the full matrix $\{\mathbf{v}_c\}_{c\in\mathcal{T}}$ is computed once and reused, so the SID-level term costs a single matrix product at inference time.

\textbf{Inference.} At inference, the start point is an all-mask sequence $[\mathbb{m}, \cdots,\mathbb{m}]$. A single forward pass of the encoder-decoder then produces both the $L\times M$ table of average generation probabilities $\bar{p}_{1|0,1}(\cdot|x_0,\mathcal{H},\theta)$ and the user-side representation $\mathbf{u}$. To ensure that the generated SID corresponds to a valid item, we perform trie-constrained generation over $\mathcal{T}$, \textit{i.e.}, the set of valid trajectories in the SID prefix tree~\citep{hokamp2017lexically, hu2026empowering}, following many existing GR works~\citep{yang2025sparse,hu2026empowering,hu2026ids,jin2025generative,wang2026one,guo2026mlps,hou2025generating}. Formally, the generation score of SID $c$ aligns with the training process:
\begin{equation}
\label{eq:inference}
\mathrm{softmax}\Big(\sum\nolimits_{l=1}^{L}\log\bar{p}_{1|0,1}(c^l|x_0,\mathcal{H},\theta)
\Big)+\mathrm{softmax}\big(\langle \mathbf{u},\mathbf{v}_c\rangle\big),
\end{equation}
where softmax is taken over $\mathcal{T}$, mirroring the two levels of
Eq.~\ref{eq:dual-level-loss}. The $K$ valid SIDs with the highest scores are returned as the recommendation list.

\section{Experiments}
\subsection{Experimental Setup}
\textbf{Datasets and Evaluation.} We evaluate all methods on eight datasets spanning multiple categories: three from Amazon Reviews 2014 \citep{mcauley2015image} (Sports, Beauty, Toys), four from Amazon Reviews 2023 \citep{hou2026bridging} (Instruments, Scientific, Games, Arts), and Yelp \citep{wang2024learnable}.  Dataset statistics are provided in Table \ref{tab:dataset_stats}. We report Recall@$K$ (R@$K$) and NDCG@$K$ (N@$K$) over the entire item catalog
with $K\in\{5,10\}$, following common practice \citep{rajput2023recommender, hou2025generating}. To evaluate model efficiency, we report the per-epoch training time and the total inference time on the whole test set.

\textbf{Baselines.} We compare SPRINT against representative and state-of-the-art methods from three families: \textbf{(1) Item ID-based:} GRU4Rec \citep{hidasi2016session}, SASRec \citep{kang2018self}, BERT4Rec \citep{sun2019bert4rec}, PreferDiff \citep{liu2025preference}, FAVE \citep{Shi2026fave}; \textbf{(2) Semantic ID-based (autoregressive):} TIGER \citep{rajput2023recommender}, ActionPiece \citep{Hou2025Action}, COBRA \citep{yang2025sparse}, SID-MLP \citep{guo2026mlps}; \textbf{(3) Semantic ID-based (non-autoregressive):} RPG \citep{hou2025generating}, LLaDA-Rec \citep{shi2025llada}, DiffGRM \citep{liu2026diffgrm}, OneGR \citep{wang2026one}. Detailed descriptions of these methods are provided in Appendix \ref{sec:baseline}.

\textbf{Implementation Details.} Each item is tokenized into $L{=}4$ tokens and the maximum length of user history is fixed at $20$. The backbone is a bidirectional Transformer following prior work \citep{liu2026diffgrm} with a $1$-layer encoder and a $4$-layer decoder. $\rho$ is searched over \{0, 0.25, 0.5, 0.75, 1\}. The number of negative SIDs sampled in Eq. \ref{eq:dual-level-loss} is set to $100$ on all datasets except Arts, where it is set to $1000$. All results are averaged over five runs with different random seeds. Additional implementation details are provided in Appendix \ref{sec:Implementation Details}.

\begin{table*}[t]
\centering
\setlength{\tabcolsep}{2.5pt}
\renewcommand{\arraystretch}{0.9}
\caption{Performance comparison between baselines and our proposed method. The best performance is in \textbf{bold} and the second-best performance is \underline{underlined}. ``*'' denotes the improvement over the second-best performance is statistically significant ($p<0.05$) according to a paired t-test. Due to space limitations, the results of N@10 and R@10 are provided in Table \ref{tab:main_results_at10}. Results on Yelp and Arts datasets are provided in Table \ref{tab:more_datasets}.} 
\vspace{-1em}
\label{tab:main_results}
\resizebox{\textwidth}{!}{%
\begin{tabular}{l cc cc cc cc cc cc}
\toprule
\multirow{2}{*}{Method}
& \multicolumn{2}{c}{Sports} & \multicolumn{2}{c}{Beauty}
& \multicolumn{2}{c}{Toys} & \multicolumn{2}{c}{Scientific}
& \multicolumn{2}{c}{Instruments} & \multicolumn{2}{c}{Games} \\
\cmidrule(lr){2-3}\cmidrule(lr){4-5}\cmidrule(lr){6-7}\cmidrule(lr){8-9}\cmidrule(lr){10-11}\cmidrule(lr){12-13}
& R@5 & N@5 & R@5 & N@5 & R@5 & N@5 & R@5 & N@5 & R@5 & N@5 & R@5 & N@5 \\
\midrule
\multicolumn{13}{c}{\textit{Item ID-based}} \\
\midrule
GRU4Rec     & 0.0129 & 0.0086 & 0.0164 & 0.0099 & 0.0097 & 0.0059 & 0.0184 & 0.0128 & 0.0297 & 0.0196 & 0.0461 & 0.0307 \\
SASRec      & 0.0233 & 0.0154 & 0.0387 & 0.0249 & 0.0463 & 0.0306 & 0.0240 & 0.0152 & 0.0331 & 0.0211 & 0.0516 & 0.0323 \\
BERT4Rec    & 0.0115 & 0.0075 & 0.0203 & 0.0124 & 0.0116 & 0.0071 & 0.0157 & 0.0100 & 0.0255 & 0.0160 & 0.0315 & 0.0199 \\
PreferDiff  & 0.0275 & 0.0190 & 0.0455 & 0.0317 & 0.0603 & 0.0403 & 0.0161 & 0.0114 & 0.0211 & 0.0139 & 0.0356 & 0.0241 \\
FAVE        & 0.0302 & 0.0194 & 0.0585 & 0.0404 & 0.0616 & 0.0444 & 0.0227 & 0.0146 & 0.0327 & 0.0209 & 0.0545 & 0.0351 \\
\midrule
\multicolumn{13}{c}{\textit{Semantic ID-based (autoregressive)}} \\
\midrule
TIGER       & 0.0264 & 0.0181 & 0.0454 & 0.0321 & 0.0521 & 0.0371 & 0.0282 & 0.0183 & 0.0359 & 0.0233 & 0.0529 & 0.0348 \\
ActionPiece & 0.0316 & 0.0205 & 0.0511 & 0.0340 & 0.0477 & 0.0315 & 0.0272 & 0.0174 & 0.0388 & 0.0247 & 0.0585 & 0.0382 \\
COBRA       & 0.0305 & 0.0215 & 0.0537 & 0.0395 & 0.0619 & 0.0462 & 0.0272 & 0.0179 & 0.0344 & 0.0231 & 0.0477 & 0.0307 \\
SID-MLP     & 0.0265 & 0.0176 & 0.0440 & 0.0302 & 0.0375 & 0.0242 & 0.0295 & 0.0192 & 0.0395 & 0.0257 & 0.0578 & 0.0378 \\
\midrule
\multicolumn{13}{c}{\textit{Semantic ID-based (non-autoregressive)}} \\
\midrule
RPG         & 0.0314 & 0.0216 & 0.0550 & 0.0381 & 0.0592 & 0.0401 & 0.0257 & 0.0174 & 0.0362 & 0.0241 & 0.0579 & 0.0397 \\
LLaDA-Rec   & 0.0281 & 0.0184 & 0.0506 & 0.0352 & 0.0506 & 0.0344 & 0.0310 & 0.0203 & 0.0406 & 0.0268 & \underline{0.0623} & \underline{0.0415} \\
DiffGRM     & 0.0363 & 0.0245 & 0.0603 & 0.0414 & 0.0618 & 0.0455 & \underline{0.0405} & \underline{0.0273} & \underline{0.0426} & \underline{0.0281} & 0.0617 & 0.0408 \\
OneGR       & \underline{0.0377} & \underline{0.0257} & \underline{0.0628} & \underline{0.0426} & \underline{0.0635} & \underline{0.0464} & 0.0357 & 0.0245 & 0.0368 & 0.0249 & 0.0544 & 0.0369 \\
\midrule
\textbf{SPRINT} & \textbf{0.0411$^*$} & \textbf{0.0270$^*$} & \textbf{0.0675$^*$} & \textbf{0.0451$^*$} & \textbf{0.0708$^*$} & \textbf{0.0501$^*$} & \textbf{0.0444$^*$} & \textbf{0.0301$^*$} & \textbf{0.0462$^*$} & \textbf{0.0305$^*$} & \textbf{0.0636$^*$} & \textbf{0.0427$^*$} \\
\textit{Improv.} & 9.02\% & 5.06\% & 7.48\% & 5.87\% & 11.50\% & 7.97\% & 9.63\% & 10.26\% & 8.45\% & 8.54\% & 2.09\% & 2.89\% \\
\bottomrule
\end{tabular}%
}
\end{table*}

\vspace{-0.5em}
\begin{table*}[t]
\centering
\setlength{\tabcolsep}{2.5pt}
\renewcommand{\arraystretch}{0.9}
\caption{The inference (\textit{i.e.}, generation) time cost (s) comparison and the speedup of our method SPRINT over baselines. A comparison of training time cost is provided in Table \ref{tab:training_speedup}. Besides, the comparison between SPRINT and other efficiency-oriented methods is provided in Table \ref{tab:inference_extra}.}
\label{tab:inference_speedup}
\vspace{-1em}
\resizebox{\textwidth}{!}{%
\begin{tabular}{l cc cc cc cc cc cc}
\toprule
\multirow{2}{*}{Methods}
& \multicolumn{2}{c}{Sports} & \multicolumn{2}{c}{Beauty} & \multicolumn{2}{c}{Toys}
& \multicolumn{2}{c}{Scientific} & \multicolumn{2}{c}{Instruments} & \multicolumn{2}{c}{Games} \\
\cmidrule(lr){2-3}\cmidrule(lr){4-5}\cmidrule(lr){6-7}\cmidrule(lr){8-9}\cmidrule(lr){10-11}\cmidrule(lr){12-13}
& Cost & \textcolor{gray!150}{Speedup} & Cost & \textcolor{gray!150}{Speedup} & Cost & \textcolor{gray!150}{Speedup} & Cost & \textcolor{gray!150}{Speedup} & Cost & \textcolor{gray!150}{Speedup} & Cost & \textcolor{gray!150}{Speedup} \\
\midrule
TIGER       & 80.91  & \textcolor{gray!150}{30.53$\times$}  & 39.69  & \textcolor{gray!150}{26.11$\times$}  & 32.45  & \textcolor{gray!150}{25.35$\times$}  & 150.82 & \textcolor{gray!150}{36.97$\times$}  & 166.92 & \textcolor{gray!150}{36.13$\times$}  & 280.00 & \textcolor{gray!150}{37.09$\times$} \\
ActionPiece & 126.36 & \textcolor{gray!150}{47.68$\times$}  & 86.16  & \textcolor{gray!150}{56.68$\times$}  & 71.46  & \textcolor{gray!150}{55.83$\times$}  & 185.67 & \textcolor{gray!150}{45.51$\times$}  & 234.12 & \textcolor{gray!150}{50.67$\times$}  & 385.05 & \textcolor{gray!150}{51.00$\times$} \\
\midrule
RPG         & \underline{24.70}  & \textcolor{gray!150}{9.32$\times$}   & \underline{14.79}  & \textcolor{gray!150}{9.73$\times$}   & \underline{12.85}  & \textcolor{gray!150}{10.04$\times$}  & \underline{34.25}  & \textcolor{gray!150}{8.39$\times$}   & \underline{40.18}  & \textcolor{gray!150}{8.70$\times$}   & \underline{64.43}  & \textcolor{gray!150}{8.53$\times$} \\
LLaDA-Rec   & 286.60 & \textcolor{gray!150}{108.15$\times$} & 183.70 & \textcolor{gray!150}{120.86$\times$} & 161.20 & \textcolor{gray!150}{125.94$\times$} & 413.70 & \textcolor{gray!150}{101.40$\times$} & 468.40 & \textcolor{gray!150}{101.39$\times$} & 756.80 & \textcolor{gray!150}{100.24$\times$} \\
DiffGRM     & 37.97  & \textcolor{gray!150}{14.33$\times$}  & 23.60  & \textcolor{gray!150}{15.53$\times$}  & 20.12  & \textcolor{gray!150}{15.72$\times$}  & 53.23  & \textcolor{gray!150}{13.05$\times$}  & 59.84  & \textcolor{gray!150}{12.95$\times$}  & 96.23  & \textcolor{gray!150}{12.75$\times$} \\
OneGR       & 35.51  & \textcolor{gray!150}{13.40$\times$}  & 22.05  & \textcolor{gray!150}{14.51$\times$}  & 19.26  & \textcolor{gray!150}{15.05$\times$}  & 50.22  & \textcolor{gray!150}{12.31$\times$}  & 57.65  & \textcolor{gray!150}{12.48$\times$}  & 140.94 & \textcolor{gray!150}{18.67$\times$} \\
\textbf{SPRINT}        & \textbf{2.65}   & \textcolor{gray!150}{-}              & \textbf{1.52}   & \textcolor{gray!150}{-}              & \textbf{1.28}   & \textcolor{gray!150}{-}              & \textbf{4.08}   & \textcolor{gray!150}{-}              & \textbf{4.62}   & \textcolor{gray!150}{-}              & \textbf{7.55}   & \textcolor{gray!150}{-} \\
\bottomrule
\end{tabular}%
}
\vspace{-1.5em}
\end{table*}

\subsection{Main results}
\textbf{Recommendation accuracy.} As shown in Table~\ref{tab:main_results}, SPRINT outperforms all baselines on every dataset, improving N@5 by up to $10.26\%$ and R@5 by up to $11.50\%$ over the second-best method. Two factors account for this. First, the dual-level flow contrastive objective addresses three requirements at once: per-token supervision, negative preference modeling, and inter-token coherence modeling, none of which a purely per-token objective can satisfy on its own. Second, single-step generation scores an SID as a whole rather than committing to tokens one at a time, which avoids the error accumulation inherent to token-by-token decoding, where an early mistaken token constrains every subsequent one \citep{wang2026one}.

\textbf{Generation efficiency.} We compare inference-time generation cost of the whole test dataset under identical settings (batch size, beam width where applicable, and sequence length; details are provided in Appendix \ref{sec:Implementation Details}) between SPRINT and accuracy-oriented multi-step GR methods (comparison with efficiency-oriented methods is provided in Table \ref{tab:inference_extra}). As shown in Table~\ref{tab:inference_speedup}, SPRINT is the fastest method on all six datasets. Against RPG, the second-fastest baseline, it delivers speedups of $8.39\times$ to $10.04\times$; against AR methods such as TIGER and ActionPiece the gap widens to $25.35\times$--$56.68\times$, and against the multi-step masked-diffusion method LLaDA-Rec it reaches up to $125.94\times$. This advantage follows directly from single-step generation: SPRINT performs one forward pass per user regardless of SID length $L$, whereas AR and NAR baselines generally pay $L$ passes, further multiplied by the beam width when a ranked list is required. SPRINT also outperforms other single-step methods (RPG and OneGR) due to our lightweight and concise design.

\subsection{Ablation Study}
\label{sec:ablation}
\begin{table*}[t]
\centering
\setlength{\tabcolsep}{4pt}
\renewcommand{\arraystretch}{0.9}
\caption{Ablation study on six datasets.}
\label{tab:ablation_at10}
\vspace{-1em}
\resizebox{\textwidth}{!}{%
\begin{tabular}{l cc cc cc cc cc cc}
\toprule
\multirow{2}{*}{Methods}
& \multicolumn{2}{c}{Sports} & \multicolumn{2}{c}{Beauty} & \multicolumn{2}{c}{Toys}
& \multicolumn{2}{c}{Scientific} & \multicolumn{2}{c}{Instruments} & \multicolumn{2}{c}{Games} \\
\cmidrule(lr){2-3}\cmidrule(lr){4-5}\cmidrule(lr){6-7}\cmidrule(lr){8-9}\cmidrule(lr){10-11}\cmidrule(lr){12-13}
& R@5 & N@5 & R@5 & N@5 & R@5 & N@5 & R@5 & N@5 & R@5 & N@5 & R@5 & N@5 \\
\midrule
\textbf{SPRINT (full)} & \textbf{0.0411} & \textbf{0.0270} & \textbf{0.0675} & \textbf{0.0451} & \textbf{0.0708} & \textbf{0.0501} & \textbf{0.0444} & \textbf{0.0301} & \textbf{0.0462} & \textbf{0.0305} & \textbf{0.0636} & \textbf{0.0427} \\
\midrule
OPQ $\rightarrow$ RQ-VAE    & 0.0284 & 0.0186 & 0.0505 & 0.0348 & 0.0452 & 0.0307 & 0.0277 & 0.0176 & 0.0383 & 0.0250 & 0.0585 & 0.0385 \\
OPQ $\rightarrow$ RQ-Kmeans & 0.0296 & 0.0195 & 0.0518 & 0.0352 & 0.0546 & 0.0355 & 0.0308 & 0.0202 & 0.0401 & 0.0262 & 0.0593 & 0.0390 \\
\midrule
w/o Token-level flow & 0.0387 & 0.0256 & 0.0639 & 0.0417 & 0.0679 & 0.0486 & 0.0428 & 0.0294 & 0.0454 & 0.0294 & 0.0624 & 0.0403 \\
w/o SID-level flow   & 0.0297 & 0.0197 & 0.0540 & 0.0366 & 0.0650 & 0.0470 & 0.0313 & 0.0214 & 0.0317 & 0.0210 & 0.0452 & 0.0297 \\
w/o Contrast         & 0.0250 & 0.0176 & 0.0533 & 0.0375 & 0.0613 & 0.0453 & 0.0258 & 0.0190 & 0.0255 & 0.0179 & 0.0399 & 0.0282 \\
\bottomrule
\end{tabular}%
}
\vspace{-1.5em}
\end{table*}

To assess the contribution of each component in SPRINT, we conduct ablation studies along two axes in Table~\ref{tab:ablation_at10}. Overall, the full model achieves the best performance across all six datasets and both metrics, confirming that every component contributes positively.

We first replace OPQ tokenizer with RQ-VAE and RQ-Kmeans. Both alternatives lead to a substantial performance drop across all datasets. This indicates that the residual, sequentially dependent codes produced by RQ-based tokenizers are ill-suited to our single-step formulation. The parallel, mutually independent subspaces of OPQ align naturally with single-step generation, where all tokens of an SID are predicted simultaneously rather than conditioned on one another. This is why we adopt OPQ following other NAR models \citep{liu2026diffgrm, hou2025generating, wang2026one}.

We then ablate three key ingredients of our model. Removing the token-level flow (\textit{i.e.}, the first term in Eq.~\ref{eq:dual-level-loss} and Eq.~\ref{eq:inference}) consistently degrades accuracy, as the model loses the fine-grained supervision that sharpens the prediction at each position. Removing the SID-level flow (\textit{i.e.}, the second term in Eq.~\ref{eq:dual-level-loss} and Eq.~\ref{eq:inference}) causes a much sharper drop: without it, an item is scored as a product of independent per-position beliefs, so positions that are each individually plausible may combine into a tuple that indexes an inaccurate SID. This is exactly the reason why we design the SID-level term to restore token coherence. The gap between the two variants is itself informative. The SID-level term scores the $L$ tokens jointly as a single item, a signal that no amount of per-position supervision can supply, whereas the per-position resolution it lacks is partially recovered by the joint score itself, since $\mathbf{v}_c$ is assembled from the token embeddings of all positions. Further replacing the final objective in Eq. \ref{eq:dual-level-loss} with Eq.~\ref{eq:ce} (\emph{w/o contrast}) causes an even larger drop, since this variant discards two signals at once: the inter-token dependency, as above, and the contrastive normalization over non-target items. The objective is then driven purely by the likelihood of the ground-truth SID and never learns what separates the target item from competing candidates, which is precisely what ranking quality depends on.

\subsection{Hyperparameter Analysis}

\begin{wrapfigure}{r}{0.6\columnwidth}
\vspace{-4.5em}
  \includegraphics[width=0.6\columnwidth]{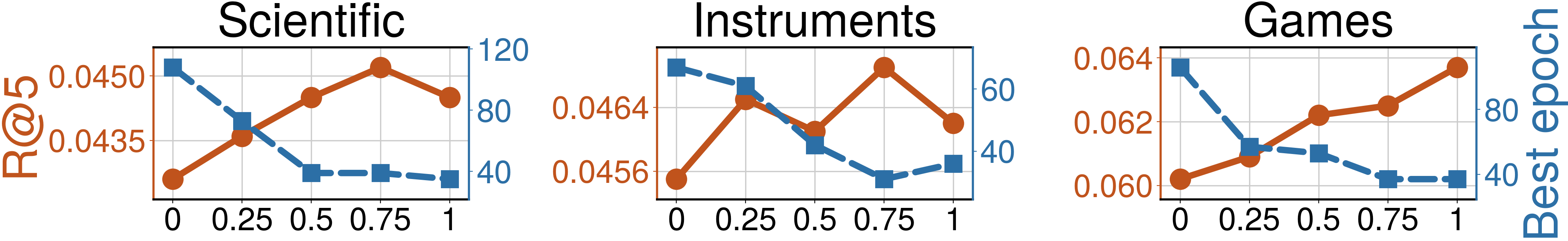}
  \vspace{-1.5em}
  \captionof{figure}{Effect of $\rho$ on model performance and the epoch to achieve the best performance.}
  \label{fig:rho}

    \includegraphics[width=0.6\columnwidth]{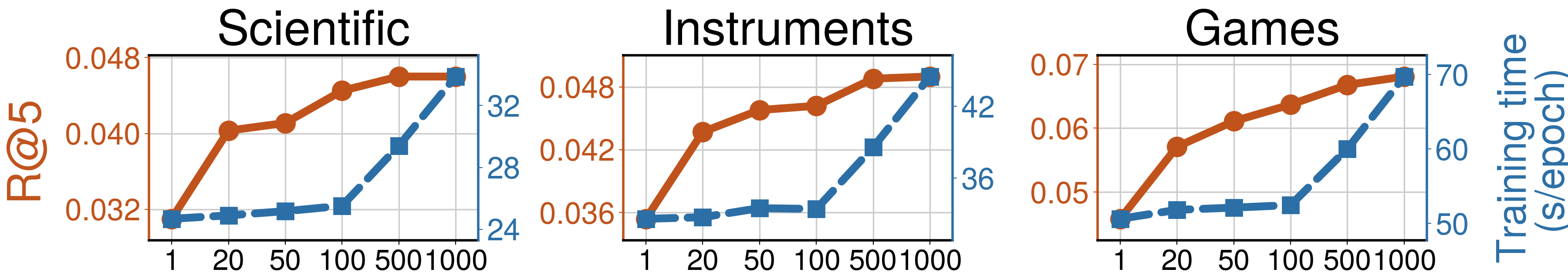}
    \vspace{-1.5em}
  \captionof{figure}{Effect of $N$ on model performance and the per-epoch training time cost.}
  \label{fig:N}
  \vspace{-1em}
\end{wrapfigure}

\textbf{Effect of $\rho$.} The ratio $\rho$ controls the fraction of training samples anchored at the all-mask source $r=0$, while the rest start from partially committed sources. As shown in Figure~\ref{fig:rho}, accuracy generally improves as $\rho$ grows and peaks at $\rho=0.75$ or $1$. Convergence also accelerates as $\rho$ grows. Samples with $r>0$ may teach the model to complete a partly decided SID, which differs from the inference condition. Raising $\rho$ redirects this gradient signal to the all-mask source that the model actually decodes from, which speeds up convergence.

\textbf{Effect of $N$.} $N$ is the number of negative SIDs sampled in Eq.~\ref{eq:dual-level-loss}. As shown in Figure~\ref{fig:N}, accuracy rises steadily as $N$ grows to $100$ and then improves only slowly up to $1000$. In contrast, training time stays nearly flat up to $N=100$ but grows sharply afterwards. More negatives help the model separate the target from competing SIDs, but beyond $100$ they bring little new signal at a much higher cost. We therefore set $N=100$ by default. In practice, $N$ can be adjusted to balance accuracy and training cost according to the needs of a specific system.

\section{Conclusion}
Existing GR models generally generate an item token by token, which causes high inference latency that hinders their deployment in real-world RSs. In this paper, we proposed SPRINT, a single-step generative recommendation model built from a new perspective of average probability velocity. We view SID generation as a flow of token generation probabilities from an all-mask sequence to the target SID. Instead of following the instantaneous velocity of this flow step by step, we characterize it by its average velocity over the whole generation process. We prove that this average velocity is fully determined by the average generation probability of each token at each position. Single-step generation thus reduces to generating the probabilities of all tokens simultaneously in one forward pass. To restore the inter-token coherence lost in this process, we design a dual-level flow contrastive objective. It contrasts the target SID against negative SIDs at both the token and SID levels. Experiments on 8 real-world datasets show that SPRINT significantly improves both generation efficiency and recommendation accuracy. The limitations of our method are discussed in Appendix~\ref{sec:limitations}.

\subsection*{AI use statement}
In this work, we used LLMs to polish the language of the manuscript, \textit{e.g.}, correcting grammar, improving readability, and refining wording. We additionally employed LLM agents to automate parts of our experimental workflow, \textit{e.g.}, orchestrating batched hyperparameter runs and collecting the resulting metrics. LLMs were not involved in any substantive aspect of the research: the research idea, methodology and model design. All AI-assisted text and results were reviewed and verified by the authors, who take full responsibility for the content of this paper.

\subsection*{Ethics statement}
This paper aims to develop a single-step generative recommendation model to improve the generation efficiency of GR models. We confirm that we do not anticipate any negative impacts and our work does not violate the ICLR code of ethics.

\subsection*{Reproducibility statement}
All results reported in our paper are fully reproducible. The pseudocode of our proposed method is provided in Algorithms \ref{alg:training} and \ref{alg:inference}. The implementation details are provided in Appendix \ref{sec:Implementation Details} and the hyperparameter settings of our method are provided in Table \ref{tab:hyperparams}. The code of our method is available at \url{https://github.com/iamZhuoCai/SPRINT}.

\bibliography{iclr2027_conference}
\bibliographystyle{iclr2027_conference}

\newpage
\appendix
\renewcommand{\theequation}{\thesection.\arabic{equation}}
\begin{center}
    \LARGE \bf {Appendix of SPRINT}
\end{center}
\etocdepthtag.toc{mtappendix}
\etocsettagdepth{mtchapter}{none}
\etocsettagdepth{mtappendix}{subsection}
\tableofcontents

\newpage
\section{More Related Work}
\label{sec:more_related_work}
\subsection{Generative Recommendation}
\subsubsection{Autoregressive Generative Recommendation.}
Autoregressive (AR) GR decodes the tokens of an SID from left to right, imposing a strict sequential dependency \citep{rajput2023recommender, Hou2025Action, yang2025sparse, fang2026prism, qiao2026text, ju2025generative, he2026reasoning, Wang2026gflowgr, hou2026expressiveness}. TIGER~\citep{rajput2023recommender} pioneered this paradigm by quantizing item embeddings into semantic IDs (SIDs) via residual quantization~\citep{lee2022autoregressive, kong2025minionerec, penha2025semantic, hou2023learning} and decoding them
token-by-token. 

These methods derives its SIDs purely from content embeddings; however, the resulting token space reflects semantic similarity alone: two items that users frequently co-consume need not be close in it, so the identifiers themselves carry no collaborative signal. To overcome this issue, some works incorporate collaborative signals pre-trained recommendation models into item tokenization stage \citep{wang2024learnable, liu2026best, wang2025empowering, ye2025dual, li2026unigrec}. LETTER \citep{wang2024learnable} resolves this by extending the RQ-VAE tokenizer with collaborative alignment and a diversity regularizer that additionally mitigates code assignment imbalance, EAGER \citep{wang2024eager} by keeping behavioral and semantic information in two separate streams so that neither is collapsed into the other, and LC-Rec \citep{zheng2024adapting} by integrating collaborative semantics into an LLM-based recommender.

These methods enrich what the tokenizer encodes, but they all learn it in a stage that precedes and is separate from the recommender, so the identifiers are frozen before the recommender ever sees them and can never adapt to the recommendation objective. To address this limitation, a plenty of works explore to develop end-to-end GR \citep{liu2025generative, fu2026differentiable, li2026unigrec, wang2026pit, jiang2026end, bai2026bi, hu2026once}. ETEGRec \citep{liu2025generative} eliminates this decoupling by optimizing tokenizer and recommender end to end, using recommendation-oriented alignment together with an alternating optimization scheme to keep joint training stable. A residual form of the same rigidity is that an item receives the same SID regardless of the actions surrounding it; ActionPiece \citep{Hou2025Action} addresses this by tokenizing action sequences contextually, so that an item's tokens depend on the context in which it occurs.

Inspired by the success of large reasoning models \citep{tang2026think}, another line of work equips GR with reasoning ability. One branch performs explicit reasoning in natural language. OneRec-Think \citep{liu2025onerec} aligns itemic tokens with the textual space of an LLM. It then activates reasoning through reasoning scaffolding and enhances it with a recommendation-specific reward. SIDReasoner \citep{he2026reasoning} first strengthens the alignment between SIDs and language. It then optimizes reasoning trajectories via outcome-driven reinforcement learning, without explicit reasoning annotations. The other branch reasons implicitly in latent space. S$^2$GR \citep{guo2026s2gr} performs stepwise latent reasoning guided by the semantics of hierarchical SID codes. LASAR \citep{chen2026lasar} adopts recurrent hidden-state feedback and adapts the reasoning depth to each sample.

Even with well-constructed identifiers, generation itself remains error-prone: an SID is a tuple drawn from $L$ codebooks, and an unconstrained decoder can emit combinations that correspond to no item in the corpus. Trie-constrained decoding answers this by building a prefix tree over all valid SIDs and masking, at every
step, any token that would leave a valid root-to-leaf path \citep{su2026vectorizing, wang2026one, li2024survey, hu2026ids}.

What none of these advances changes is the decoding paradigm itself. Tokens are still emitted strictly left to right, so an SID of $L$ tokens costs $L$ forward passes, further multiplied by the beam width when a ranked list is required, and this cost is inherent to autoregression rather than to any particular instantiation of it. Relaxing the left-to-right constraint is precisely what the non-autoregressive methods reviewed next set out to do.

\subsubsection{Non-autoregressive Generative Recommendation}
The $L$ sequential forward passes that autoregression imposes are a consequence of its factorization, not of the recommendation task, which motivates non-autoregressive (NAR) decoding: tokens can be emitted in parallel from any position without left-to-right constraint \citep{ren2024non}. Removing the causal factorization, however, removes with it the conditioning that kept the tokens of an SID coherent, and the NAR literature is largely an account of how that lost dependency is put back.

The dominant answer is to restore it iteratively. Following the rapid progress of diffusion models \citep{ho2020denoising, yang2023generate, cai2025unleashing, cai2026steering, qu2026diffusion} and diffusion language models \citep{nie2026large, gulrajani2023likelihood, sahoo2024simple}, diffusion-based GR \citep{qu2026diffusion, shi2025llada, liu2026diffusion, zhang2025gdiffmae, zhu2026time} reveals the positions of an SID over several rounds, each round conditioning on the tokens already committed.  LLaDA-Rec \citep{shi2025llada} and MaskGR \citep{shah2025masked} adopt discrete diffusion to decode in arbitrary rather than left-to-right order, and subsequent work refines which positions to reveal and when, for instance through history-aware mask position allocation \citep{mu2026masked} or on-policy coherent noising \citep{liu2026diffgrm}. Order is thus no longer fixed, but the number of network evaluations is not reduced: competitive accuracy still requires token-by-token refinement \citep{shi2025llada}, so $L$ forward passes remain, and collapsing the schedule to a single step causes accuracy to collapse with it (Table~\ref{tab:pilot_experiments}).

A second answer avoids iteration by making a single parallel prediction expressive enough to stand on its own. RPG \citep{hou2025generating, xia2026unleash} predicts all positions simultaneously under a multi-token objective and excludes incoherent outputs through graph-constrained decoding. The dependency, however, is enforced at decoding time rather than learned, and expressiveness is bought by lengthening the SID to $16$--$64$ tokens, which enlarges the embedding table to roughly $4\times$--$16\times$ that of the $4$-token SIDs used by most NAR models \citep{shi2025llada, liu2026diffgrm, mu2026masked}. OneGR \citep{wang2026one} selects tokens by an A$^*$ search \citep{hart1968formal} that weighs each prefix against a look-ahead bound over the remaining positions, so decoding is not strictly single-pass in scoring but remains sequential and multiple steps.

These methods either still require multiple steps, substantially enlarge the embedding table, or rely on time-consuming search without a principled single-step guarantee. To bridge these gaps, in this paper, we propose a theoretically principled, effective and efficient single-step generative recommendation model.

\subsection{Flow matching}
Flow matching \citep{Lipman2023flow, Liu2023flow, jin2025pyramidal, dao2023flow} was introduced to shorten the long sampling trajectories of diffusion models, learning a time-dependent velocity field that transports a prior to the data distribution along substantially straighter paths. What it learns, however, is the \emph{instantaneous} velocity, which describes an infinitesimal move and must therefore be integrated: sampling still solves the generation trajectory over many network evaluations, so the cost is reduced rather than removed.

Single-step flow matching addresses this residual cost by changing the object of learning \citep{luo2026soflow, geng2025mean, geng2026improved}. Geng et al. \citep{geng2025mean} model the \emph{average} velocity over a finite interval, which spans the trajectory in one evaluation, and recovers it through an identity that relates the average field to the instantaneous one by differentiating with respect to the state; later work improves its training stability \citep{geng2026improved} or attains one-step transport by learning straight trajectories directly \citep{kornilov2024optimal}. This resolves the step count, but the enabling identity is defined only where the state is continuous and differentiable, and has no counterpart over the categorical token space in which SIDs live.

Discrete flow matching \citep{Campbell2024generative,  gat2024discrete} supplies the missing half of the picture, extending the framework to categorical data by formulating generation as a continuous-time Markov chain over discrete states. Yet it remains built on instantaneous transition rates, and therefore inherits exactly the limitation that motivated the average-velocity view in the first place: sampling is again iterative. The two lines are thus complementary but disjoint, one achieving single-step generation only in continuous space, the
other achieving discreteness only at multi-step cost.

Within recommendation, flow matching has so far been used to model preference transport rather than to accelerate generation, whether for sequential recommendation \citep{Shi2026fave, Liu2025flow}, and multi-domain settings \citep{ye2026gaussian}, but none of these works operates in discrete SID generation domain or generates an item in a single step. We close the gap between the two lines by deriving the average probability velocity directly in discrete token space.

\section{Notations}
The descriptions of the notations used in this paper are presented in Table \ref{tab:notation}.
\begin{table}[t]
\caption{Notations and explanations.}
\label{tab:notation}
\centering
\small
\begin{tabular}{@{}ll@{}}
\toprule
\textbf{Notation} & \textbf{Explanation} \\
\midrule
$\mathcal{H}$ & Discrete token sequence of the user's interacted items \\
$n$ & Maximum length of items in the user history \\
$K$ & Number of items in the returned recommendation list \\
$L$ & Number of positions in an SID \\
$M$ & Size of each codebook \\
$\mathcal{C}^l$ & Codebook of position $l$, from which the $l$-th token of an SID is drawn \\
$c=[c^1,\dots,c^L]$ & Semantic ID (SID) of an item; $c^l$ is its token at position $l$ \\
$\mathcal{T}$ & Set of valid trajectories in the SID prefix tree \\
$\mathbb{m}$ & Mask token, denoting a position that has not yet been decided \\
\midrule
$x_0$ & Source of the flow, \textit{i.e.}, the all-mask sequence $[\mathbb{m},\dots,\mathbb{m}]$ \\
$x_1$ & Target of the flow, \textit{i.e.}, the SID of the target item $i_{n+1}$ \\
$x_r$ & State of the SID being generated at time $r$ \\
$z$ & Generic current state of a position \\
$r$, $t$ & Start and terminal times of the interval over which the velocity is averaged \\
$\tau$ & Integration variable ranging over $[r,t]$ \\
$p_t(x)$ & Time-indexed probability path over SIDs \\
$\pi(x_0,x_1)$ & Coupling of source and target endpoints \\
$\kappa_t$ & Scheduler measuring how much of the target SID has been decided \\
$\dot\kappa_t$ & Time derivative of $\kappa_t$, \textit{i.e.}, the instantaneous switching rate \\
$\delta_y(\cdot)$ & Indicator that a position (or an SID) equals $y$ \\
\midrule
$v^l_\tau(x^l,z)$ & Instantaneous probability velocity: rate at which position $l$ jumps from $z^l$ to $x^l$ \\
$\bar v^l_{r,t}(x^l,z)$ & Average probability velocity of position $l$ over the interval $[r,t]$ \\
$\lambda_{r,t}$ & Interval-level scalar rate, $\tfrac{1}{t-r}\ln\tfrac{1-\kappa_r}{1-\kappa_t}$ \\
$p_{1|\tau}(x^l|z)$ & Instantaneous target posterior at position $l$ \\
$\bar p_{1|r,t}(x^l|z,\mathcal{H},\theta)$ & Average generation probability learned by the network; the token predictor \\
\midrule
$\theta$ & Parameters of the bidirectional Transformer (1-layer encoder, 4-layer decoder) \\
$\mathbf{h}^l$ & Last-layer decoder hidden state at SID position $l$ \\
$\mathbf{e}_{c^l}$ & Embedding of token $c^l$ in codebook $\mathcal{C}^l$ \\
$\mathbf{u}$ & User/Sequence-side joint representation, obtained from $[\mathbf{h}^1,\dots,\mathbf{h}^L]$ \\
$\mathbf{v}_c$ & SID-side joint representation of a candidate SID $c$ \\
$\langle\cdot,\cdot\rangle$ & Cosine similarity \\
$N$ & Number of negative SID samples \\
$\mathcal{D}$ & Training set of (history, target SID) pairs \\
$\mathcal{L}$ & Dual-level flow contrastive loss \\
\bottomrule
\end{tabular}
\end{table}

\section{Theoretical Derivations}
\subsection{Derivation of the interpolant in Equation \ref{eq:inst}}
\label{derivation_interpolant}
The marginal velocity $v_\tau^l(x^l,z)$ is intractable, so we first derive the velocity conditioned on a fixed target $x_1$ (the source $x_0$ can be omitted since it is an all-mask sequence), for which the path is a simple interpolant. We use the convex-interpolant conditional path, whose $\tau$-derivative follows since only $\kappa_\tau$ depends on $\tau$:
\begin{equation}
\label{eq:cond-path}
    p_\tau(x^l|x_1)=(1-\kappa_\tau)\delta_{\mathbb{m}}(x^l)+\kappa_\tau\delta_{x_1}(x^l),
\end{equation}
\begin{equation}
    \label{p_derivative_tau1}
    \dot{p}_\tau(x^l|x_1)=\dot{\kappa}_\tau[\delta_{x_1}(x^l)-\delta_\mathbb{m}(x^l)].
\end{equation}

A velocity is admissible only if it generates this path through the Kolmogorov equation \citep{Campbell2024generative, gat2024discrete}, \textit{i.e.}, the rate of change of the probability at $x^l$ equals the net probability flowing into $x^l$ from every current state $z^l$:
\begin{equation}
    \label{p_derivative_tau2}
    \dot{p}_\tau(x^l|x_1)=\sum_{z^l}v_\tau^l(x^l,z^l|x_1)p_\tau(z^l|x_1).
\end{equation}

\textbf{Proposition 1.} \textit{The conditional velocity generating Eq. \ref{eq:cond-path} in denoiser mode is:}
\begin{equation}
\label{eq:cond-vel}
    v_\tau^l(x^l,z|x_1)=\frac{\dot{\kappa}_\tau}{1-\kappa_\tau}[\delta_{x_1}(x^l)-\delta_{z}(x^l)].
\end{equation}

\textbf{Proof.} In denoiser mode (predicting the clean target, as in $x$-prediction diffusion) the source coefficient is zero, leaving a two term ansatz supported on the target and the current state:
\begin{equation}
\label{v_a1_a2}
    v_\tau^l(x^l,z|x_1)=a_\tau^1\delta_{x_1}(x^l)+a_\tau^2\delta_{z}(x^l).
\end{equation}

Substitute Eq. \ref{v_a1_a2} into the right-hand side of Eq. \ref{p_derivative_tau2}, we have:
\begin{align}
  \sum_{z^l}v^l_\tau(x^l,z^l\mid x_1)\,p_\tau(z^l\mid x_1)
    &=a^1_\tau\,\delta_{x_1}(x^l)
        \underbrace{\sum_{z^l}p_\tau(z^l\mid x_1)}_{=\,1}
      +a^2_\tau
        \underbrace{\sum_{z^l}\delta_{z^l}(x^l)\,p_\tau(z^l\mid x_1)}
                   _{=\,p_\tau(x^l\mid x_1)}
        \nonumber\\
    &=a^1_\tau\,\delta_{x_1}(x^l)+a^2_\tau\,p_\tau(x^l\mid x_1)
        \nonumber\\
    &=a^1_\tau\,\delta_{x_1}(x^l)
      +a^2_\tau\big[(1-\kappa_\tau)\delta_{\mathbb{m}}(x^l)+\kappa_\tau\delta_{x_1}(x^l)\big]
        \nonumber\\
    &=(a^1_\tau+a^2_\tau\kappa_\tau)\,\delta_{x_1}(x^l)
      +a^2_\tau(1-\kappa_\tau)\,\delta_{\mathbb{m}}(x^l).
      \label{eq:rhs-expanded}
\end{align}

Setting \eqref{eq:rhs-expanded} equal to \eqref{p_derivative_tau1} and matching the coefficients of the linearly independent indicators
$\delta_{x_1}$ and $\delta_\mathbb{m}$ (distinct since $x_1\neq\mathbb{m}$):
\begin{equation}
  a^1_\tau+a^2_\tau\kappa_\tau=\dot\kappa_\tau,
  \qquad
  a^2_\tau(1-\kappa_\tau)=-\dot\kappa_\tau.
  \label{eq:matching}
\end{equation}

Solving the second for $a^2_\tau$ and back-substituting:
\begin{equation}
  a^2_\tau=-\frac{\dot\kappa_\tau}{1-\kappa_\tau},
  \qquad
  a^1_\tau=\dot\kappa_\tau-a^2_\tau\kappa_\tau
    =\frac{\dot\kappa_\tau(1-\kappa_\tau)+\dot\kappa_\tau\kappa_\tau}{1-\kappa_\tau}
    =\frac{\dot\kappa_\tau}{1-\kappa_\tau}.
  \label{eq:solved}
\end{equation}
Substituting \eqref{eq:solved} into \eqref{v_a1_a2} gives
\eqref{eq:cond-vel}. Proof completed.

\subsection{Validity of the Conditional Velocity in Proposition 1}
\label{sec:validity}

For \eqref{eq:cond-vel} to be a legitimate CTMC generator \citep{norris1998markov, Campbell2024generative}, two conditions must hold.

\paragraph{Condition 1 (columns sum to zero --- probability conservation).}
Mass leaving $z$ must reappear elsewhere, so summing over all destinations
$x^l$ must give zero:
\begin{equation}
  \sum_{x^l}v^l_\tau(x^l,z\mid x_1)
    =\frac{\dot\kappa_\tau}{1-\kappa_\tau}
      \Big[\sum_{x^l}\delta_{x_1}(x^l)-\sum_{x^l}\delta_z(x^l)\Big]
    =\frac{\dot\kappa_\tau}{1-\kappa_\tau}\,[1-1]=0.
  \label{eq:cond1}
\end{equation}

\paragraph{Condition 2 (non-negative off-diagonal --- valid jump rates).}
For $x^l\neq z$ we have $\delta_z(x^l)=0$, so we have,
\begin{equation}
  v^l_\tau(x^l,z\mid x_1)
    =\frac{\dot\kappa_\tau}{1-\kappa_\tau}\,\delta_{x_1}(x^l)\ge 0,
  \label{eq:cond2}
\end{equation}
using $\dot\kappa_\tau\ge0$ (monotone scheduler) and $1-\kappa_\tau\ge0$.
Both hold, so \eqref{eq:cond-vel} is a valid velocity generating
\eqref{eq:cond-path}.

\subsection{Conditional Average Probability Velocity}
\label{sec:con_ave_pro_vel}
To jump over a whole interval $[r,t]$ in one evaluation instead of many small steps, we use the time-averaged velocity:
\begin{equation}
\label{eq:int_v}
    \bar v^l_{r,t}(x^l,z)=\frac{1}{t-r}\int_r^t v^l_\tau(x^l,z)\,d\tau.
\end{equation}

\textbf{Proposition 2.} \textit{The conditional average velocity retains the same ``target minus current" structure as the instantaneous one, with the time-varying rate $\tfrac{\dot\kappa_\tau}{1-\kappa_\tau}$ collapsed into a single interval-level scalar $\lambda_{r,t}$:}
\begin{equation}
\label{eq:prop2}
    \bar v^l_{r,t}(x^l,z\mid x_1)=\lambda_{r,t}\big[\delta_{x_1}(x^l)-\delta_z(x^l)\big],
\qquad
\lambda_{r,t}:=\frac{1}{t-r}\ln\frac{1-\kappa_r}{1-\kappa_t},
\end{equation}
\textit{where coefficient $\lambda_{r,t}$ is the time-averaged switching rate over $[r,t]$: it is the total rate accumulated across the interval, $\int_r^t\tfrac{\dot\kappa_\tau}{1-\kappa_\tau}d\tau=\ln\tfrac{1-\kappa_r}{1-\kappa_t}$, divided by the interval length $t-r$. Because the direction of transport $[\delta_{x_1}-\delta_z]$ does not change with $\tau$, averaging the velocity reduces entirely to averaging this scalar rate—which is exactly what makes a single interval-spanning step well-defined.}

\textbf{Proof.} Substitute \ref{eq:cond-vel} into \ref{eq:int_v}, we have:
\begin{equation}
\begin{aligned}
        \bar{v}_{r,t}^l(x^l,z|x_1)&=\frac{1}{t-r}\int_{r}^t v_\tau^l(x^l,z|x_1)d\tau \\
        &=\frac{1}{t-r}\int_r^t\frac{\dot{\kappa_\tau}}{1-\kappa_\tau}[\delta_{x_1}(x^l)-\delta_z(x^l)]d\tau \\
        &=\frac{1}{t-r}[\delta_{x_1}(x^l)-\delta_z(x^l)]\int_r^t\frac{\dot{\kappa}_\tau}{1-\kappa_\tau}d\tau \\
        &=\frac{1}{t-r}[\delta_{x_1}(x^l)-\delta_z(x^l)]\left[-\ln(1-\kappa_\tau)\right]_r^t \\
        &=\frac{1}{t-r}\ln\frac{1-\kappa_r}{1-\kappa_t}[\delta_{x_1}(x^l)-\delta_z(x^l)] \\
        &=\lambda_{r,t}[\delta_{x_1}(x^l)-\delta_z(x^l)],
\end{aligned}
\end{equation}
where $\lambda_{r,t}=\frac{1}{t-r}\ln\frac{1-\kappa_r}{1-\kappa_t}$. Proof completed.

\subsection{Marginal Average Probability Velocity}
\label{sec:marginal-average-vel}
We need the marginal velocity, since the target $x_1$ is unknown at generation time. By the standard flow-matching marginalization, the marginal velocity that generates $p_\tau$ is the average of the conditional velocities weighted by the target posterior $p_\tau(x_1\mid z)$—the model's belief, given the current partial state $z$, about which item is being generated:
\begin{equation}
\label{eq:instan-vel}
    v^l_\tau(x^l,z)=\sum_{x_1}v^l_\tau(x^l,z\mid x_1)\,p_\tau(x_1\mid z).
\end{equation}

The interval-average of this quantity is subtle for one reason: the posterior $p_\tau(x_1\mid z)$ depends on $\tau$ and lives inside the time integral $[r,t]$. Consequently the marginal average velocity is not obtained by taking the conditional result \ref{eq:prop2} and replacing $x_1$ by a single posterior mean; it requires a specifically weighted average posterior, defined next.

\textbf{Proposition 3.} \textit{Let $p_{1\mid\tau}(x^l\mid z)$ be the instantaneous target posterior and $\bar p_{1\mid r,t}(x^l\mid z)$ is $\tfrac{\dot\kappa_\tau}{1-\kappa_\tau}$-weighted time average over $[r,t]$:}
\begin{equation}
\label{eq:prop3}
    p_{1\mid\tau}(x^l\mid z):=\sum_{x_1}\delta_{x_1}(x^l)\,p_\tau(x_1\mid z),
\qquad
\bar p_{1\mid r,t}(x^l\mid z):=\frac{\int_r^t\frac{\dot\kappa_\tau}{1-\kappa_\tau}\,p_{1\mid\tau}(x^l\mid z)\,d\tau}{\int_r^t\frac{\dot\kappa_\tau}{1-\kappa_\tau}\,d\tau}.
\end{equation}

\textit{Then the marginal average velocity keeps the same ``posterior minus current" form as \ref{eq:prop2}, with the weighted-average posterior in place of the target indicator:}
\begin{equation}
\label{eq:instan-average-vel}
    \bar v^l_{r,t}(x^l,z)=\lambda_{r,t}\big[\bar p_{1\mid r,t}(x^l\mid z)-\delta_z(x^l)\big].
\end{equation}

\textbf{Proof.} Substitute the conditional velocity \ref{eq:cond-vel} into the marginalization \ref{eq:instan-vel}, we have:
\begin{equation}
\label{eq:derivation-instan-vel}
    \begin{aligned}
    v^l_\tau(x^l,z)&=\sum_{x_1}v^l_\tau(x^l,z\mid x_1)\,p_\tau(x_1\mid z)\\
&=\frac{\dot\kappa_\tau}{1-\kappa_\tau}\sum_{x_1}\big[\delta_{x_1}(x^l)-\delta_z(x^l)\big]\,p_\tau(x_1\mid z)\\
&=\frac{\dot\kappa_\tau}{1-\kappa_\tau}\Big[\underbrace{\sum_{x_1}\delta_{x_1}(x^l)\,p_\tau(x_1\mid z)}_{=\,p_{1\mid\tau}(x^l\mid z)}-\delta_z(x^l)\underbrace{\sum_{x_1}p_\tau(x_1\mid z)}_{=\,1}\Big]\\
&=\frac{\dot\kappa_\tau}{1-\kappa_\tau}\big[p_{1\mid\tau}(x^l\mid z)-\delta_z(x^l)\big].
\end{aligned}
\end{equation}

The first sum is the posterior by definition in \ref{eq:prop3} following \citep{gat2024discrete}. $\sum_{x_1}p_\tau(x_1|z)=1$ by the law of total probability.

Insert Eq. \ref{eq:derivation-instan-vel} into the Eq. \ref{eq:int_v} of the average velocity and separate the posterior term from the current-state term:
\begin{equation}
    \bar v^l_{r,t}(x^l,z)
=\underbrace{\frac{1}{t-r}\int_r^t\frac{\dot\kappa_\tau}{1-\kappa_\tau}\,p_{1\mid\tau}(x^l\mid z)\,d\tau}_{(\mathrm{I})}
\;-\;\delta_z(x^l)\,\underbrace{\frac{1}{t-r}\int_r^t\frac{\dot\kappa_\tau}{1-\kappa_\tau}\,d\tau}_{(\mathrm{II})}.
\end{equation}

Term $(\mathrm{II})$ is exactly $\lambda_{r,t}$ by Eq. \ref{eq:prop2}. For term $(\mathrm{I})$, multiply and divide by the normalizer $\int_r^t\frac{\dot\kappa_\tau}{1-\kappa_\tau}d\tau=(t-r)\lambda_{r,t}$, which turns the integral into the weighted-average posterior in \ref{eq:prop3}:
\begin{equation}
    \begin{aligned}
(\mathrm{I})
&=\frac{1}{t-r}\int_r^t\frac{\dot\kappa_\tau}{1-\kappa_\tau}\,p_{1\mid\tau}(x^l\mid z)\,d\tau\\[4pt]
&=\frac{1}{t-r}\left(\int_r^t\frac{\dot\kappa_\tau}{1-\kappa_\tau}\,d\tau\right)\cdot\frac{\displaystyle\int_r^t\frac{\dot\kappa_\tau}{1-\kappa_\tau}\,p_{1\mid\tau}(x^l\mid z)\,d\tau}{\displaystyle\int_r^t\frac{\dot\kappa_\tau}{1-\kappa_\tau}\,d\tau}\\
&=\frac{1}{t-r}\left(\int_r^t\frac{\dot\kappa_\tau}{1-\kappa_\tau}\,d\tau\right)\bar p_{1\mid r,t}(x^l\mid z)\\
&=\frac{1}{t-r}\,(t-r)\,\lambda_{r,t}\,\bar p_{1\mid r,t}(x^l\mid z)\\
&=\lambda_{r,t}\,\bar p_{1\mid r,t}(x^l\mid z).
\end{aligned}
\end{equation}

Substituting $(\mathrm{I})$ and $(\mathrm{II})$ back,
$$
\bar v^l_{r,t}(x^l,z)=\lambda_{r,t}\,\bar p_{1\mid r,t}(x^l\mid z)-\lambda_{r,t}\,\delta_z(x^l)=\lambda_{r,t}\big[\bar p_{1\mid r,t}(x^l\mid z)-\delta_z(x^l)\big],
$$
which is \ref{eq:instan-average-vel}. Proof completed.

\section{More Experiments}
\subsection{Datasets Statistics}
We evaluate all methods on eight datasets spanning multiple categories: three from Amazon Reviews 2014 \citep{mcauley2015image} (Sports, Beauty, Toys)\footnote{https://cseweb.ucsd.edu/~jmcauley/datasets/amazon/links.html}, four from Amazon Reviews 2023 \citep{hou2026bridging} (Instruments, Scientific, Games, Arts)\footnote{https://huggingface.co/datasets/McAuley-Lab/Amazon-Reviews-2023}, and Yelp\footnote{https://business.yelp.com/data/resources/open-dataset/} \citep{wang2024learnable}. The statistics of our used datasets are reported in Table \ref{tab:dataset_stats}. Each user's review history is arranged into a single sequence ordered by timestamp. For all methods, the commonly used leave-one-out strategy is adopted for data splitting, following previous works \citep{kang2018self, rajput2023recommender}: in each sequence, the most recent interaction is used for testing, the second most recent for validation, and the remainder for training. 

\begin{table}[t]
\centering
\caption{Statistics of the datasets used in this paper.}
\label{tab:dataset_stats}
\begin{tabular}{lcccc}
\toprule
\textbf{Dataset} & \textbf{\#Users} & \textbf{\#Items} & \textbf{\#Interactions} & \textbf{Avg.\ Sequence Length} \\
\midrule
\textbf{Sports}      & 35,598 & 18,357 & 296,337 & 8.32 \\
\textbf{Beauty}      & 22,363 & 12,101 & 198,502 & 8.87 \\
\textbf{Toys}        & 19,412 & 11,924 & 167,597 & 8.63 \\
\textbf{Instruments} & 57,439 & 24,587 & 511,836 & 8.91 \\
\textbf{Scientific}  & 50,985 & 25,848 & 412,947 & 8.10 \\
\textbf{Games}       & 94,762 & 25,612 & 814,586 & 8.60 \\
\textbf{Yelp}       & 30,431 & 20,033 & 316,354 & 10.40 \\
\textbf{Arts}       & 197,287 & 89,959 & 1,786,437 & 9.06 \\
\bottomrule
\end{tabular}
\end{table}

\subsection{Baseline Description}
\label{sec:baseline}
We compare SPRINT against representative and SOTA baselines spanning three families.

\paragraph{Item ID-based Methods.}
\begin{itemize}[left=0pt]
    \item \textbf{GRU4Rec}~\citep{hidasi2016session} pioneers RNN-based session recommendation, employing GRU units to encode the sequence of intra-session interactions for next-item prediction.

    \item \textbf{SASRec}~\citep{kang2018self} adopts self-attention encoder that adaptively weighs previously interacted items to capture sequential dynamics. This attains the long-range modeling capacity of RNNs while retaining the parallel training efficiency of Transformers.

    \item \textbf{BERT4Rec}~\citep{sun2019bert4rec} replaces left-to-right decoding with a deep bidirectional Transformer trained under a Cloze (masked-item) objective.

    \item \textbf{PreferDiff}~\citep{liu2025preference} is a diffusion-based recommender that recasts the BPR objective into a log-likelihood ranking loss over multiple negative samples, optimized via a variational upper bound with a cosine (rather than MSE) reconstruction error.

    \item \textbf{FAVE}~\citep{Shi2026fave} is a flow matching-based recommendation model that learns a direct user-to-item trajectory via an average-velocity formulation, thereby avoiding the prior mismatch and iterative-solver redundancy of the noise-to-data paradigm.
\end{itemize}

\paragraph{Semantic ID-based Methods (Autoregressive).}
\begin{itemize}[left=0pt]
    \item \textbf{TIGER}~\citep{rajput2023recommender} assigns each item a semantic ID produced by an RQ-VAE over content embeddings, and trains a Transformer to autoregressively generate the next item's semantic ID.

    \item \textbf{ActionPiece}~\citep{Hou2025Action} performs context-aware tokenization of action sequences, merging co-occurring feature patterns in the spirit of BPE so that identical actions receive different tokens depending on their surrounding context. The resulting context-sensitive vocabulary improves autoregressive generation over context-agnostic SID tokenizers.

    \item \textbf{COBRA}~\citep{yang2025sparse} unifies sparse and dense retrieval through cascaded generation: it first produces a sparse semantic ID and then, conditioned on it, a dense vector, both refined end-to-end in a coarse-to-fine manner. At inference, a BeamFusion strategy combines beam search with nearest-neighbor scores to balance accuracy and diversity.

    \item \textbf{SID-MLP}~\citep{guo2026mlps} proposes an MLP-centric distillation framework for efficient GR inference. It retains one-layer attention module to obtain the first token and then uses cascaded MLPs to generate the remaining tokens sequentially.
\end{itemize}

\paragraph{Semantic ID-based Methods (Non-autoregressive).}
\begin{itemize}[left=0pt]
    \item \textbf{RPG}~\citep{hou2025generating} generates all tokens of a long, unordered semantic ID in parallel within a single decoding step, decoupling the number of decoding passes from ID length.

    \item \textbf{LLaDA-Rec}~\citep{shi2025llada} casts SID generation as discrete masked diffusion, denoising SID tokens in parallel and in a flexible order rather than strictly left-to-right. This bidirectional, any-order formulation better respects the joint structure of the tokens that collectively identify an item.

    \item \textbf{DiffGRM}~\citep{liu2026diffgrm} replaces the autoregressive decoder with a masked discrete diffusion model, enabling bidirectional context and any-order parallel generation of SID tokens. It further introduces parallel semantic encoding, on-policy coherent noising that concentrates supervision on hard-to-predict tokens, and confidence-guided parallel denoising for diverse top-$K$ candidate generation.

    \item \textbf{OneGR}~\citep{wang2026one} produces position-wise scores in one pass, and then adopt A$^*$ search over valid SIDs to retrieval top-$K$ valid SIDs.
\end{itemize}

\begin{table}[t]
\centering
\small
\setlength{\tabcolsep}{5pt}
\caption{Hyperparameter settings across datasets.}
\label{tab:hyperparams}
\begin{tabular}{lcccccccc}
\toprule
\textbf{Hyperparameter} & \textbf{Sports} & \textbf{Beauty} & \textbf{Toys} & \textbf{Scientific} & \textbf{Instruments} & \textbf{Games} & \textbf{Yelp} & \textbf{Arts} \\
\midrule
learning rate      & 1e-4  & 1e-4   & 1e-4  & 1e-4    & 1e-4    & 1e-4  & 1e-4 & 1e-4 \\
warmup steps       & 10,000 & 10,000 & 10,000 & 10,000     & 10,000     & 10,000 & 10,000  & 10,000  \\
dropout rate       & 0.1    & 0.1    & 0.1    & 0.1     & 0.1     & 0.1  & 0.2 & 0.2   \\
weight decay & 0    & 0    & 0    & 0     & 0     & 0  & 0 & 0.05 \\
$d_m$               & 256    & 256    & 256   & 256     & 256     & 256  & 512 & 512   \\
$d_{ff}$               & 1024   & 1024   & 1024   & 1024     & 1024     & 1024 & 2048 & 1024   \\
num\_heads          & 4     & 4 & 4      & 4      & 4     & 4     & 4 & 8    \\
$L$                 & 4      & 4  &4    & 4      & 4 & 4     & 4 & 4     \\
$M$                 & 256    & 256  &256  & 256    & 256     & 256     & 256 & 256     \\
encoder layers     & 1      & 1   & 1   & 1      & 1     & 1     & 1   & 1  \\
decoder layers & 4      & 4      & 4  &  4   & 4     & 4     & 4 & 4     \\
$\rho$            & 1    & 1    & 1   & 0.75     & 0.75     & 1  & 1 & 1  \\
num\_neg\_samples & 100 & 100 & 100 & 100 & 100 & 100 & 100 & 1000 \\
max\_seq\_length  & 20     & 20     & 20     & 20     & 20     & 20  & 20 & 20  \\
max\_epochs         & 150    & 150    & 150    & 150     & 150     & 150  & 150 & 150   \\
early stop patience & 10   & 10     & 10     & 10     & 10     & 10  & 10 & 10  \\
\bottomrule
\end{tabular}
\end{table}

\subsection{More Implementation Details}
\label{sec:Implementation Details}
We employ $L=4$ codebooks, each containing $M=256$ codewords (\textit{i.e.}, tokens) following existing non-autoregressive GR models \citep{shi2025llada, liu2026diffgrm, wang2026one}. This means each item is discretized into $L=4$ tokens. The tokens are derived using OPQ following previous work \citep{liu2026diffgrm, wang2026one}. In the bidirectional Transformer, the token embedding dimension is $d_{m}=256$ following previous work \citep{shi2025llada, li2023diffurec}. The numbers of encoder layer and decoder layer are 1 and 4, respectively. The feed-forward layer dimension $d_{ff}$ is 1024 and the number of attention heads is 4. The MLP to derive $\mathbf{u}$ and $\mathbf{v}_c$ is a two layer architecture (\textit{i.e.}, $1024\to256\to\text{ReLU}\to256$). Following \citep{liu2026diffgrm}, all 4 tokens of an item are mapped into a 256-dimension embedding before being fed into the encoder. Training/Validation/Test batch sizes are set to 1024/256/256. The maximum length of user history is fixed at $n=20$. For fair comparison, we use \texttt{sentence-t5-base} \citep{ni2022sentence} as the semantic encoder to derive items' content embeddings in the tokenization stage following prior work \citep{liu2026diffgrm, wang2026one, hou2025generating}, so that no method benefits from a stronger semantic encoder. Model parameters are optimized using AdamW \citep{loshchilov2017decoupled}, with the learning rate searched over $\{1\mathrm{e}{-4},5\mathrm{e}{-4},1\mathrm{e}{-3},5\mathrm{e}{-3},1\mathrm{e}{-2}\}$. The maximum number of training epochs is set to 150. Following previous work \citep{liu2026diffgrm}, we use early stopping with patience 10 to stop training if the validation score $0.8\times \text{N}@10 +0.2\times \text{R}@10$ does not improve for 10 consecutive epochs. The hyperparameter values used in our implementation are reported in Table \ref{tab:hyperparams}.

\textbf{Settings for efficiency comparison.} For fair comparison of model efficiency, we set all methods' training/val/test batch sizes to 1024/256/256, maximum sequence length to 20, beam width to 10 if applicable, token embedding size to 256. Other settings are set according to the best setting specific to datasets in their original paper. These settings are only used for efficiency comparison. In accuracy comparison, these hyperparameters follow the original papers and are tuned per dataset.

\subsection{More Comparison Results ($K=10$)}
\label{sec:more_comparison}

We report the performance comparison of R@10 and N@10 in Table \ref{tab:main_results_at10}. Similar to the overall comparison reported in Table \ref{tab:main_results}, our proposed model SPRINT also consistently outperform baseline methods in terms of R@10 and N@10. 

\begin{table*}[t]
\centering
\setlength{\tabcolsep}{2.5pt}
\caption{Performance comparison between baselines and our proposed method. The best performance is in \textbf{bold} and the second-best performance is \underline{underlined}. ``*'' denotes the improvement over the second-best performance is statistically significant ($p<0.05$) according to a paired t-test.}
\label{tab:main_results_at10}
\resizebox{\textwidth}{!}{%
\begin{tabular}{l cc cc cc cc cc cc}
\toprule
\multirow{2}{*}{Method}
& \multicolumn{2}{c}{Sports} & \multicolumn{2}{c}{Beauty}
& \multicolumn{2}{c}{Toys} & \multicolumn{2}{c}{Scientific}
& \multicolumn{2}{c}{Instruments} & \multicolumn{2}{c}{Games} \\
\cmidrule(lr){2-3}\cmidrule(lr){4-5}\cmidrule(lr){6-7}\cmidrule(lr){8-9}\cmidrule(lr){10-11}\cmidrule(lr){12-13}
& R@10 & N@10 & R@10 & N@10 & R@10 & N@10 & R@10 & N@10 & R@10 & N@10 & R@10 & N@10 \\
\midrule
\multicolumn{13}{c}{\textit{Item ID-based}} \\
\midrule
GRU4Rec     & 0.0204 & 0.0110 & 0.0283 & 0.0137 & 0.0176 & 0.0084 & 0.0272 & 0.0156 & 0.0453 & 0.0246 & 0.0712 & 0.0387 \\
SASRec      & 0.0350 & 0.0192 & 0.0605 & 0.0318 & 0.0675 & 0.0374 & 0.0379 & 0.0197 & 0.0525 & 0.0273 & 0.0823 & 0.0421 \\
BERT4Rec    & 0.0191 & 0.0099 & 0.0347 & 0.0170 & 0.0203 & 0.0099 & 0.0264 & 0.0134 & 0.0412 & 0.0211 & 0.0530 & 0.0267 \\
PreferDiff  & 0.0405 & 0.0218 & 0.0660 & 0.0388 & 0.0851 & 0.0483 & 0.0216 & 0.0132 & 0.0294 & 0.0166 & 0.0501 & 0.0287 \\
FAVE        & 0.0455 & 0.0244 & 0.0847 & 0.0489 & 0.0855 & 0.0520 & 0.0359 & 0.0188 & 0.0518 & 0.0271 & 0.0858 & 0.0451 \\
\midrule
\multicolumn{13}{c}{\textit{Semantic ID-based (autoregressive)}} \\
\midrule
TIGER       & 0.0400 & 0.0225 & 0.0648 & 0.0384 & 0.0712 & 0.0432 & 0.0446 & 0.0236 & 0.0566 & 0.0300 & 0.0823 & 0.0442 \\
ActionPiece & 0.0500 & 0.0264 & 0.0775 & 0.0424 & 0.0730 & 0.0396 & 0.0437 & 0.0227 & 0.0609 & 0.0318 & 0.0910 & 0.0486 \\
COBRA       & 0.0434 & 0.0257 & 0.0725 & 0.0456 & 0.0781 & 0.0515 & 0.0424 & 0.0228 & 0.0509 & 0.0285 & 0.0750 & 0.0395 \\
SID-MLP     & 0.0417 & 0.0225 & 0.0715 & 0.0391 & 0.0600 & 0.0314 & 0.0459 & 0.0244 & 0.0612 & 0.0327 & 0.0916 & 0.0486 \\
\midrule
\multicolumn{13}{c}{\textit{Semantic ID-based (non-autoregressive)}} \\
\midrule
RPG         & 0.0463 & 0.0263 & 0.0809 & 0.0464 & 0.0869 & 0.0490 & 0.0395 & 0.0218 & 0.0545 & 0.0300 & 0.0853 & 0.0485 \\
LLaDA-Rec   & 0.0432 & 0.0233 & 0.0747 & 0.0429 & 0.0752 & 0.0423 & 0.0474 & 0.0256 & 0.0623 & 0.0337 & 0.0942 & \underline{0.0517} \\
DiffGRM     & 0.0550 & 0.0305 & 0.0876 & 0.0502 & 0.0834 & 0.0524 & \underline{0.0605} & \underline{0.0337} & \underline{0.0659} & \underline{0.0356} & \underline{0.0950} & 0.0515 \\
OneGR       & \underline{0.0574} & \underline{0.0332} & \underline{0.0914} & \underline{0.0519} & \underline{0.0874} & \underline{0.0538} & 0.0507 & 0.0294 & 0.0526 & 0.0299 & 0.0799 & 0.0451 \\
\midrule
\textbf{SPRINT} & \textbf{0.0621$^*$} & \textbf{0.0347$^*$} & \textbf{0.0987$^*$} & \textbf{0.0552$^*$} & \textbf{0.0998$^*$} & \textbf{0.0594$^*$} & \textbf{0.0656$^*$} & \textbf{0.0369$^*$} & \textbf{0.0713$^*$} & \textbf{0.0386$^*$} & \textbf{0.1001$^*$} & \textbf{0.0534$^*$} \\
\textit{Improv.} & 8.19\% & 4.52\% & 7.99\% & 6.36\% & 14.19\% & 10.41\% & 8.43\% & 9.50\% & 8.19\% & 8.43\% & 5.37\% & 3.29\% \\
\bottomrule
\end{tabular}%
}
\end{table*}

\subsection{Training Efficiency Analysis}
\label{sec:trainging_efficiency}
\begin{figure}
    \centering
    \includegraphics[width=1\linewidth]{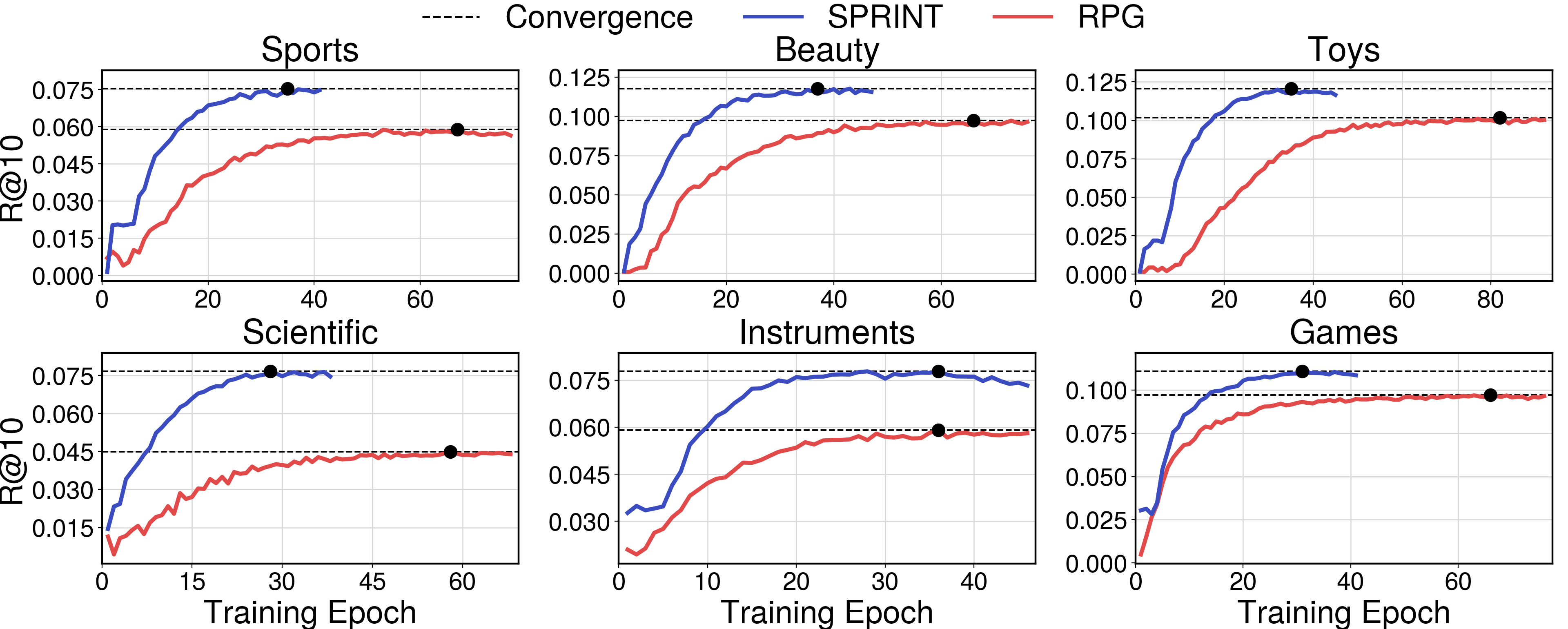}
    \caption{Training convergence comparison with RPG on six datasets.}
    \label{fig:convergence}
\end{figure}

As shown in Table \ref{tab:training_speedup}, we have the following observation:
\begin{itemize}[left=0pt]
    \item SPRINT's training time cost is also lower than that of existing AR and NAR GR models. It can be attributed to our single-step generation formulation and the lightweight model architecture, 1-layer encoder and 4-layer decoder, with no auxiliary mask-scheduling (\textit{e.g.}, history-aware mask position allocation \citep{mu2026masked}) or data-augmentation machinery (\textit{e.g.}, on-policy coherent noising \citep{liu2026diffgrm}).

    \item Besides, we also compare the convergence curves of SPRINT and the fastest baseline method RPG. As shown in Figure \ref{fig:convergence}, SPRINT converges much faster than RPG, generally requiring 20-40 epochs. Hence, this further indicates that our proposed method SPRINT can be trained efficiently.

\end{itemize}

\begin{table*}[t]
\centering
\caption{Training time per epoch (s) comparison and the speedup of our method over baselines.}
\label{tab:training_speedup}
\resizebox{\textwidth}{!}{%
\begin{tabular}{l cc cc cc cc cc cc}
\toprule
\multirow{2}{*}{Methods}
& \multicolumn{2}{c}{Sports} & \multicolumn{2}{c}{Beauty} & \multicolumn{2}{c}{Toys}
& \multicolumn{2}{c}{Scientific} & \multicolumn{2}{c}{Instruments} & \multicolumn{2}{c}{Games} \\
\cmidrule(lr){2-3}\cmidrule(lr){4-5}\cmidrule(lr){6-7}\cmidrule(lr){8-9}\cmidrule(lr){10-11}\cmidrule(lr){12-13}
& Cost & \textcolor{gray!130}{Speedup} & Cost & \textcolor{gray!130}{Speedup} & Cost & \textcolor{gray!130}{Speedup} & Cost & \textcolor{gray!130}{Speedup} & Cost & \textcolor{gray!130}{Speedup} & Cost & \textcolor{gray!130}{Speedup} \\
\midrule
TIGER       & 122.01 & \textcolor{gray!130}{6.67$\times$}  & 86.54  & \textcolor{gray!130}{6.92$\times$}  & 70.58  & \textcolor{gray!130}{6.73$\times$}  & 170.04 & \textcolor{gray!130}{6.63$\times$}  & 220.70 & \textcolor{gray!130}{6.55$\times$}  & 351.27 & \textcolor{gray!130}{6.69$\times$} \\
ActionPiece & 122.59 & \textcolor{gray!130}{6.70$\times$}  & 98.45  & \textcolor{gray!130}{7.87$\times$}  & 80.57  & \textcolor{gray!130}{7.68$\times$}  & 170.16 & \textcolor{gray!130}{6.64$\times$}  & 241.50 & \textcolor{gray!130}{7.17$\times$}  & 403.22 & \textcolor{gray!130}{7.68$\times$} \\
\midrule
RPG         & \underline{21.39}  & \textcolor{gray!130}{1.17$\times$}  & \underline{16.20}  & \textcolor{gray!130}{1.30$\times$}  & \underline{13.76}  & \textcolor{gray!130}{1.31$\times$}  & \underline{29.27}  & \textcolor{gray!130}{1.14$\times$}  & \underline{37.43}  & \textcolor{gray!130}{1.11$\times$}  & \underline{62.83}  & \textcolor{gray!130}{1.20$\times$} \\
LLaDA-Rec   & 202.50 & \textcolor{gray!130}{11.07$\times$} & 143.70 & \textcolor{gray!130}{11.49$\times$} & 118.80 & \textcolor{gray!130}{11.33$\times$} & 276.00 & \textcolor{gray!130}{10.77$\times$} & 368.90 & \textcolor{gray!130}{10.95$\times$} & 573.40 & \textcolor{gray!130}{10.92$\times$} \\
DiffGRM     & 69.33  & \textcolor{gray!130}{3.79$\times$}  & 47.29  & \textcolor{gray!130}{3.78$\times$}  & 39.03  & \textcolor{gray!130}{3.72$\times$}  & 91.36  & \textcolor{gray!130}{3.56$\times$}  & 122.52 & \textcolor{gray!130}{3.64$\times$}  & 195.26 & \textcolor{gray!130}{3.72$\times$} \\
OneGR       & 102.22 & \textcolor{gray!130}{5.59$\times$}  & 70.27  & \textcolor{gray!130}{5.62$\times$}  & 57.91  & \textcolor{gray!130}{5.52$\times$}  & 139.15 & \textcolor{gray!130}{5.43$\times$}  & 189.26 & \textcolor{gray!130}{5.62$\times$}  & 284.66 & \textcolor{gray!130}{5.42$\times$} \\
\textbf{SPRINT}        & \textbf{18.29}  & \textcolor{gray!130}{-}             & \textbf{12.51}  & \textcolor{gray!130}{-}             & \textbf{10.49}  & \textcolor{gray!130}{-}             & \textbf{25.63}  & \textcolor{gray!130}{-}             & \textbf{33.68}  & \textcolor{gray!130}{-}             & \textbf{52.49}  & \textcolor{gray!130}{-} \\
\bottomrule
\end{tabular}%
}
\end{table*}

\subsection{Comparison between SPRINT and Other Efficiency-oriented Methods}
\label{sec:other_efficiency_methods}
In addition to efficiency comparison with existing accuracy-oriented AR and NAR GR models, we also compare our model with efficiency-oriented GR models. To achieve this, we select the following method for comparison:
\begin{itemize}[left=0pt]
    \item \textbf{AtSpeed} \citep{lin2025efficient} accelerates LLM-based generative recommendation via speculative decoding. Since a ranked list requires every top-$K$ beam sequence of the target LLM to be drafted correctly, it aligns the draft model with the target model's top-$K$ sequences (AtSpeed-S), and further relaxes verification so that high-probability non-top-$K$ drafts can also be accepted (AtSpeed-R).
    \item \textbf{EARN} \citep{yang2025earn} reduces the computation and KV-cache cost of LLM-based generative recommendation. Observing that early layers exhibit dense, informative attention while later layers are largely redundant, it uses the first few layers to compress the interaction history into register tokens placed at the sequence boundaries, and processes only these register tokens in the remaining layers.
    \item \textbf{NEZHA} \citep{wang2026nezha} is a speculative decoding architecture for industrial generative recommendation that dispenses with a separate draft model and model-based verifier. It attaches a lightweight autoregressive draft head to the main model for self-drafting, and verifies drafted candidates by a hash lookup against the set of valid SIDs rather than by calling the target model.
    \item \textbf{SID-MLP} \citep{guo2026mlps} distills an autoregressive SID-based recommender (\textit{e.g.}, TIGER) into lightweight MLPs. Motivated by the observation that SID prediction becomes much easier after the first token, it freezes the Transformer encoder and replaces the attention-based decoder with prefix-conditioned MLP heads.
\end{itemize}

\begin{table}[H]
\centering
\setlength{\tabcolsep}{4pt}
\caption{Inference time cost (s) comparison across six datasets. The fastest is in \textbf{bold} and the second-fastest is \underline{underlined}.}
\label{tab:inference_extra}
\begin{tabular}{l cccccc}
\toprule
Methods & Sports & Beauty & Toys & Scientific & Instruments & Games \\
\midrule
AtSpeed  & 64.10 & 41.88 & 34.97 & 91.83  & 103.74 & 170.44 \\
EARN     & 76.07 & 47.62 & 41.91 & 109.58 & 124.79 & 204.45 \\
NEZHA    & 17.57 & 11.03 & 9.58  & 25.17  & 28.37  & 46.85  \\
SID-MLP  & \underline{5.18}  & \underline{3.18}  & \underline{2.86} & \underline{7.11} & \underline{7.67} & \underline{12.75} \\
SPRINT   & \textbf{2.65} & \textbf{1.52} & \textbf{1.28} & \textbf{4.08} & \textbf{4.62} & \textbf{7.55} \\
\bottomrule
\end{tabular}
\end{table}

For all competitive methods, we use TIGER as their backbone model. As shown Table~\ref{tab:inference_extra}, SPRINT achieves the lowest inference cost across all six datasets, requiring only $1.28$--$7.55$s to generate recommendations. AtSpeed and NEZHA accelerate autoregressive generative recommenders via speculative decoding, while EARN reduces computation by compressing the interaction history into register tokens; nevertheless, all three retain a fundamentally sequential decoding process and remain far slower than SPRINT, which is about $24$--$29\times$ faster than AtSpeed and EARN (\textit{e.g.}, $2.65$s vs. $64.10$s and $76.07$s on Sports) by predicting all tokens of an SID in a single step. Compared with NEZHA, which mitigates but does not eliminate autoregressive decoding, SPRINT is still about $6\times$ faster. The closest competitor is SID-MLP, which distills an autoregressive recommender into lightweight MLP heads; although it likewise avoids attention-based sequential decoding, SPRINT remains consistently faster (\textit{e.g.}, $1.28$s vs. $2.86$s on Toys) while delivering substantially higher recommendation accuracy (Table~\ref{tab:main_results}). These results demonstrate that the efficiency of SPRINT stems from its single-step generative formulation rather than merely from a compact architecture, making it well-suited for latency-sensitive, real-world recommendation scenarios.

\subsection{Parameter Memory}
\begin{wraptable}{r}{0.5\textwidth}
\centering
\caption{Parameter memory comparison across paradigms.}
\label{tab:params}
\begin{tabular}{@{}l l r@{}}
\toprule
Paradigms & Methods & \#Parameters \\
\midrule
\multirow{2}{*}{AR}
 & TIGER       & 13.97 M \\
 & ActionPiece & 9.58 M \\
\midrule
\multirow{4}{*}{NAR}
 & RPG         & 8.53 M -- 23.69 M \\
 & LLaDA-Rec   & 6.85 M \\
 & DiffGRM     & 5.60 M \\
 & SPRINT & 5.59 M \\
\bottomrule
\end{tabular}
\end{wraptable}

Table~\ref{tab:params} compares the parameter memory of representative generative recommenders across the autoregressive (AR) and non-autoregressive (NAR) paradigms. The vocabulary size of ActionPiece is set to 40,000 following default setting of original paper. When the SID length of RPG is set to 16, its memory is 8.53 M. However, in many datasets, RPG requires an SID length of 64, significantly increasing the memory to 23.69 M. Within these paradigms, SPRINT is the most lightweight, requiring only 5.59 M parameters, on par with the most compact baseline, DiffGRM (5.60 M), and roughly $2.5\times$ smaller than TIGER. Crucially, SPRINT attains this minimal parameter footprint while delivering the best recommendation accuracy (Table~\ref{tab:main_results}), indicating that its single-step generative formulation improves effectiveness without incurring any additional parameter cost.

\subsection{Experiments on More Diverse Datasets}
Apart from the six datasets from Amazon used in the main text, we also evaluate the effectiveness of our method on more diverse datasets, including Yelp \citep{wang2024learnable} and Arts \citep{hou2026bridging}. From the results reported in Table \ref{tab:more_datasets}, we can find that SPRINT also consistently achieves the best performance on Yelp and Arts. This verifies the effectiveness and generalization of our method.

\begin{table}[H]
\centering
\small
\caption{Performance comparison between baselines and our proposed method. The best performance is in \textbf{bold} and the second-best performance is \underline{underlined}. ``*'' denotes the improvement over the second-best performance is statistically significant ($p<0.05$) according to a paired t-test.}
\label{tab:more_datasets}
\begin{tabular}{l cccc cccc}
\toprule
\multirow{2}{*}{Methods}
 & \multicolumn{4}{c}{Yelp}
 & \multicolumn{4}{c}{Arts} \\
\cmidrule(lr){2-5} \cmidrule(lr){6-9}
 & R@5 & N@5 & R@10 & N@10
 & R@5 & N@5 & R@10 & N@10 \\
\midrule
GRU4Rec      & 0.0216 & 0.0138 & 0.0377 & 0.0190 & 0.0169 & 0.0106 & 0.0281 & 0.0142 \\
SASRec       & 0.0240 & 0.0167 & 0.0370 & 0.0209 & 0.0263 & 0.0143 & 0.0416 & 0.0192 \\
TIGER        & 0.0235 & 0.0156 & 0.0376 & 0.0202 & 0.0264 & 0.0174 & 0.0409 & 0.0220 \\
ActionPiece  & 0.0285 & 0.0182 & 0.0481 & 0.0251 & 0.0245 & 0.0165 & 0.0356 & 0.0200 \\
RPG          & 0.0216 & 0.0147 & 0.0319 & 0.0180 & 0.0246 & 0.0168 & 0.0364 & 0.0206 \\
LLaDA-Rec    & \underline{0.0322} & \underline{0.0218} & \underline{0.0502} & \underline{0.0276} & 0.0267 & 0.0173 & 0.0422 & 0.0223 \\
DiffGRM      & 0.0272 & 0.0177 & 0.0460 & 0.0237 & \underline{0.0414} & \underline{0.0291} & \underline{0.0522} & \underline{0.0326} \\
OneGR      & 0.0189 & 0.0124 & 0.0314 & 0.0164 & 0.0275 & 0.0197 & 0.0379 & 0.0231
 \\
\midrule
SPRINT & \textbf{0.0350}$^*$ & \textbf{0.0235}$^*$ & \textbf{0.0548}$^*$ & \textbf{0.0298}$^*$ & \textbf{0.0427}$^*$ & \textbf{0.0298}$^*$ & \textbf{0.0613}$^*$ & \textbf{0.0354}$^*$ \\
\textit{Improv.} & \textit{8.70\%} & \textit{7.80\%} & \textit{9.16\%} & \textit{7.97\%} & \textit{3.14\%} & \textit{2.41\%} & \textit{17.43\%} & \textit{8.59\%} \\
\bottomrule
\end{tabular}
\end{table}

\subsection{Performance of Decoder-only Architecture}
To evaluate whether SPRINT can perform well on other architectures, we additionally implement it on a decoder-only backbone (4-layer decoder) and compare it against the default encoder-decoder variant. As shown in Table~\ref{tab:decoder-only}, SPRINT delivers strong performance under both architectures. The decoder-only architecture only performs slightly worse performance than the encoder-decoder architecture on Sports, Beauty, Scientific, and Games datasets, and even achieves better performance on Toys and Instruments dataset. This demonstrates that the effectiveness of SPRINT stems from our specific model designs rather than from any particular backbone, and that the method is readily transferable across architectural choices.

\begin{table*}[t]
\centering
\caption{Comparison between encoder-decoder and decoder-only architectures for SPRINT across six datasets.}
\label{tab:decoder-only}
\resizebox{\textwidth}{!}{%
\begin{tabular}{l cccc cccc cccc}
\toprule
\multirow{2}{*}{Methods}
 & \multicolumn{4}{c}{Sports}
 & \multicolumn{4}{c}{Beauty}
 & \multicolumn{4}{c}{Toys} \\
\cmidrule(lr){2-5} \cmidrule(lr){6-9} \cmidrule(lr){10-13}
 & R@5 & N@5 & R@10 & N@10
 & R@5 & N@5 & R@10 & N@10
 & R@5 & N@5 & R@10 & N@10 \\
\midrule
SPRINT (encoder-decoder)
 & \textbf{0.0411} & \textbf{0.0270} & \textbf{0.0621} & \textbf{0.0347}
 & \textbf{0.0675} & \textbf{0.0451} & \textbf{0.0987} & \textbf{0.0552}
 & 0.0708 & 0.0501 & 0.0998 & 0.0594 \\
SPRINT (decoder-only)
 & 0.0393 & 0.0259 & 0.0611 & 0.0329
 & 0.0660 & 0.0437 & 0.0971 & 0.0537
 & \textbf{0.0739} & \textbf{0.0531} & \textbf{0.1013} & \textbf{0.0618} \\
\midrule
\multirow{2}{*}{Methods}
 & \multicolumn{4}{c}{Scientific}
 & \multicolumn{4}{c}{Instruments}
 & \multicolumn{4}{c}{Games} \\
\cmidrule(lr){2-5} \cmidrule(lr){6-9} \cmidrule(lr){10-13}
 & R@5 & N@5 & R@10 & N@10
 & R@5 & N@5 & R@10 & N@10
 & R@5 & N@5 & R@10 & N@10 \\
\midrule
SPRINT (encoder-decoder)
 & \textbf{0.0444} & \textbf{0.0301} & \textbf{0.0656} & \textbf{0.0369}
 & 0.0462 & 0.0305 & 0.0713 & 0.0386
 & 0.0636 & \textbf{0.0427} & \textbf{0.1001} & \textbf{0.0534} \\
SPRINT (decoder-only)
 & 0.0434 & 0.0294 & 0.0649 & 0.0364
 & \textbf{0.0482} & \textbf{0.0320} & \textbf{0.0730} & \textbf{0.0400}
 & \textbf{0.0639} & 0.0418 & 0.0998 & 0.0533 \\
\bottomrule
\end{tabular}%
}
\end{table*}

\subsection{More Hyperparameter Analysis}
\label{sec:more_hyper}

\begin{figure}[h]
    \centering
    \includegraphics[width=0.8\linewidth]{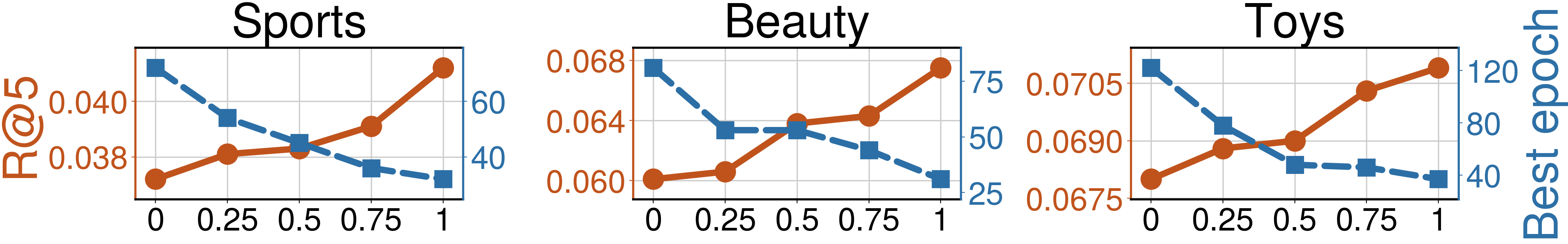}
    \caption{Effects of $\rho$ on Sports, Beauty, and Toys datasets.}
    \vspace{-1em}
    \label{fig:rho_sbt}
\end{figure}

\paragraph{Effect of $\rho$.}
We further probe the sensitivity to $\rho$ on Sports, Beauty, and Toys datasets, reporting both R@5 and the best epoch (the epoch attaining peak validation performance) in Fig.~\ref{fig:rho_sbt}. Two consistent trends emerge. First, recommendation accuracy steadily improves as $\rho$ grows and peaks at $\rho=1$ on all three datasets. This indicates that anchoring training at the all-mask source, which matches the inference condition, directly benefits single-step generation. Second, and more strikingly, $\rho$ substantially accelerates convergence. The best epoch drops sharply as $\rho$ grows, so a larger $\rho$ reaches the optimum with far fewer training epochs. Taken together, $\rho$ improves effectiveness and efficiency simultaneously. It yields better accuracy while greatly reducing the training cost, and $\rho=1$ offers the best trade-off on these three datasets.

\begin{figure}[h]
    \centering
    \includegraphics[width=0.8\linewidth]{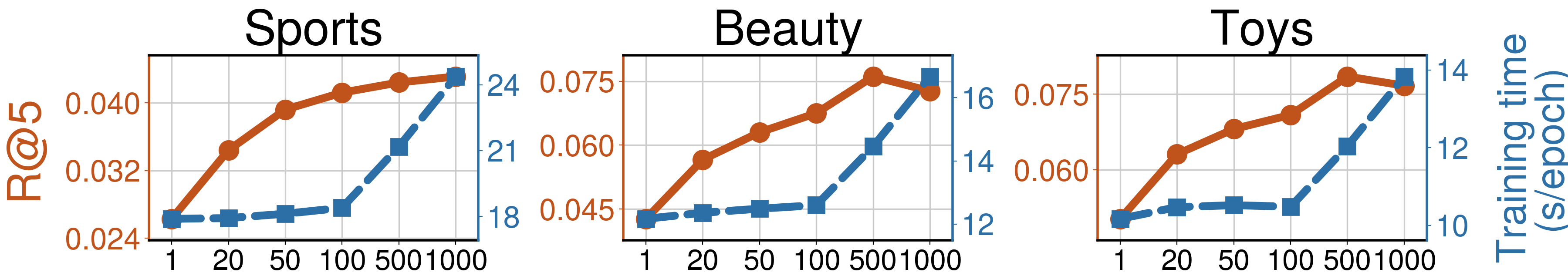}
    \caption{Effects of $N$ on Sports, Beauty, and Toys datasets.}
    \vspace{-1em}
    \label{fig:N_sbt}
\end{figure}

\paragraph{Effect of $N$.}
We further examine the effect of $N$ on Sports, Beauty, and Toys datasets, reporting both R@5 and the training time per epoch in Fig.~\ref{fig:N_sbt}. Two consistent trends emerge. First, recommendation accuracy rises steadily as $N$ grows to $100$ and then gains more slowly. Beyond $100$, R@5 keeps inching up on Sports, while it peaks at $N=500$ and slightly declines at $N=1000$ on Beauty and Toys. This suggests that a moderate number of negatives already provides most of the contrastive signal, whereas excessive negatives bring little new information and may even introduce noise. Second, the training time stays nearly flat up to $N=100$ but grows sharply afterwards, since the cost of scoring negative SIDs becomes dominant. Taken together, $N=100$ captures most of the accuracy gain at negligible extra cost, which is consistent with the observations on other datasets.


\begin{figure}
    \centering
    \includegraphics[width=1\linewidth]{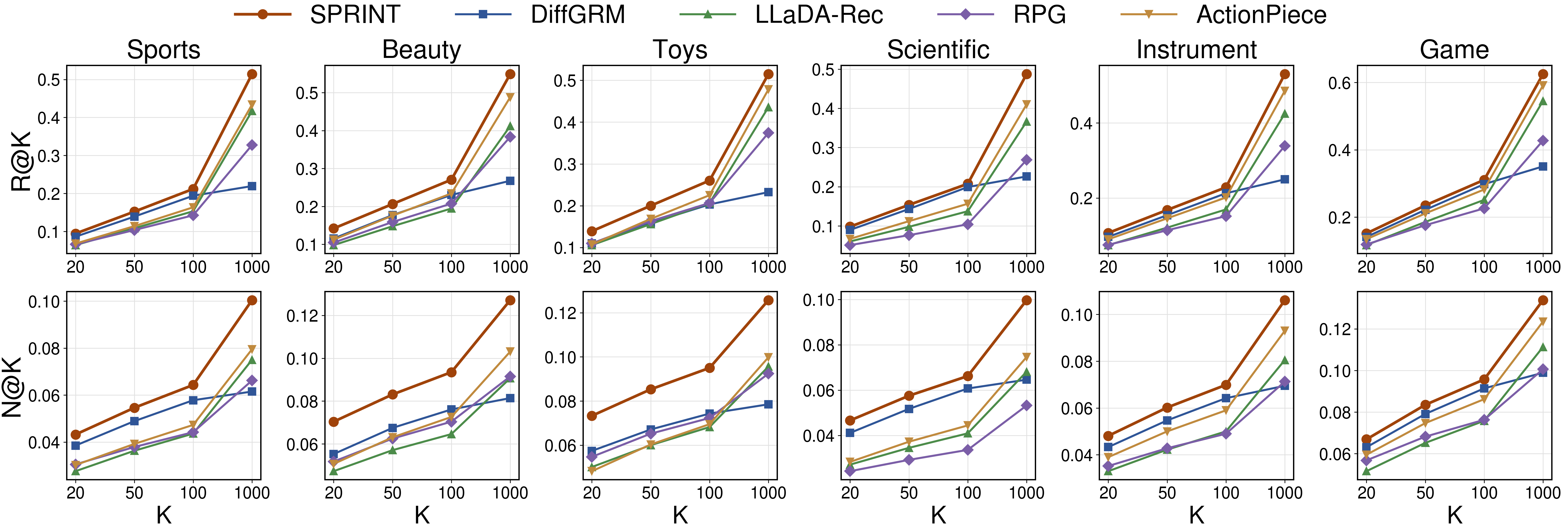}
    \caption{Performance comparison between SPRINT and other generative recommendation methods under different K values in range \{20, 50, 100, 1000\}.}
    \label{fig:large_k}
\end{figure}

\subsection{Performance Comparison Under Larger $K$}
Small cutoffs ($K\in\{5,10\}$) probe only the head of the ranking. To assess whether SPRINT also orders items well deeper into the list, Figure~\ref{fig:large_k} compares it against DiffGRM, LLaDA-Rec, RPG, and ActionPiece on R@$K$ and N@$K$ for $K\in\{20,50,100,1000\}$ across all six datasets. SPRINT remains the best method at every evaluated $K$, on every dataset, and under both metrics, and its margin over the strongest baseline does not erode as $K$ grows---indeed widening in most settings. We attribute this to SPRINT scoring each candidate SID jointly rather than committing to tokens one position at a time: joint scoring keeps the entire candidate pool well-ordered far from the top, whereas the error accumulation of token-by-token or iterative decoding degrades the tail of the ranking. This robustness at large $K$ is valuable for systems that must surface many items per user.

\subsection{Performance of TIGER using OPQ as the tokenizer}
We replace the RQ-VAE used in TIGER \citep{rajput2023recommender} with an OPQ tokenizer and report the results in Table~\ref{tab:tiger}. We observe that OPQ consistently degrades TIGER's performance across all six datasets and all metrics. This is because the left-to-right generation of TIGER inherently requires the SIDs to exhibit sequential dependencies: RQ-VAE quantizes an item embedding in a coarse-to-fine residual manner, so that each codeword is conditioned on the residual left by its predecessors, yielding a hierarchical token order that aligns naturally with the autoregressive decoding of TIGER. OPQ, in contrast, partitions the embedding into mutually independent subspaces and quantizes each in parallel; the resulting codewords carry no inherent ordering and are effectively permutation-invariant. As a result, the autoregressive decoder cannot exploit any conditional structure among the generated tokens, which explains the observed performance drop.

\begin{table*}[t]
\centering
\caption{Comparison of TIGER based on RQ-VAE and OPQ.}
\label{tab:tiger}
\resizebox{\textwidth}{!}{%
\begin{tabular}{l cccc cccc cccc}
\toprule
\multirow{2}{*}{Methods}
 & \multicolumn{4}{c}{Sports}
 & \multicolumn{4}{c}{Beauty}
 & \multicolumn{4}{c}{Toys} \\
\cmidrule(lr){2-5} \cmidrule(lr){6-9} \cmidrule(lr){10-13}
 & R@5 & N@5 & R@10 & N@10
 & R@5 & N@5 & R@10 & N@10
 & R@5 & N@5 & R@10 & N@10 \\
\midrule
TIGER (RQ-VAE)
 & \textbf{0.0264} & \textbf{0.0181} & \textbf{0.0400} & \textbf{0.0225}
 & \textbf{0.0454} & \textbf{0.0321} & \textbf{0.0648} & \textbf{0.0384}
 & \textbf{0.0521} & \textbf{0.0371} & \textbf{0.0712} & \textbf{0.0432} \\
TIGER (OPQ)
 & 0.0220 & 0.0140 & 0.0375 & 0.0190
 & 0.0390 & 0.0262 & 0.0603 & 0.0330
 & 0.0346 & 0.0223 & 0.0584 & 0.0299 \\
\midrule
\multirow{2}{*}{Methods}
 & \multicolumn{4}{c}{Scientific}
 & \multicolumn{4}{c}{Instruments}
 & \multicolumn{4}{c}{Games} \\
\cmidrule(lr){2-5} \cmidrule(lr){6-9} \cmidrule(lr){10-13}
 & R@5 & N@5 & R@10 & N@10
 & R@5 & N@5 & R@10 & N@10
 & R@5 & N@5 & R@10 & N@10 \\
\midrule
TIGER (RQ-VAE)
 & \textbf{0.0282} & \textbf{0.0183} & \textbf{0.0446} & \textbf{0.0236}
 &\textbf{ 0.0359} &\textbf{ 0.0233} & \textbf{0.0566} & \textbf{0.0300}
 & \textbf{0.0529} & \textbf{0.0348} & \textbf{0.0823} & \textbf{0.0442} \\
TIGER (OPQ)
 & 0.0257 & 0.0166 & 0.0412 & 0.0216
 & 0.0339 & 0.0219 & 0.0526 & 0.0278
 & 0.0496 & 0.0321 & 0.0794 & 0.0416 \\
\bottomrule
\end{tabular}%
}
\end{table*}

\section{Algorithms}
The algorithms of SPRINT's training and inference stage are present in Algorithms \ref{alg:training} and \ref{alg:inference}. 

\begin{figure}[t]
\begin{minipage}{\textwidth}
\begin{algorithm}[H]
\caption{Training of SPRINT}
\label{alg:training}
\begin{algorithmic}[1]
\Require Training set $\mathcal{D}=\{(\mathcal{H},x_1)\}$; sampled SID set $\mathcal{S}$;
codebooks $\{\mathcal{C}^l\}_{l=1}^{L}$
\Ensure Network parameters $\theta$
\State Tokenize all items into $L$-token SIDs with OPQ \Comment{Tokenization}
\While{not converged}
  \State Sample a mini-batch $\{(\mathcal{H},x_1)\}\subset\mathcal{D}$
  \State Obtain partially masked sequence $x_r$ according to $r$
  \State Encode $\mathcal{H}$ and decode from $L$ mask queries to obtain $\bar{p}_{1|r,1}(\cdot\,|\,x_r,\mathcal{H},\theta)\in\mathbb{R}^{L\times M}$ and $\{\mathbf{h}^l\}_{l=1}^{L}$
  \State $\mathbf{u}\leftarrow\mathrm{MLP}\big(\mathrm{Cat}(\mathbf{h}^1,\dots,\mathbf{h}^L)\big)$ \Comment{User/Sequence-side representation}
  \State $\mathbf{v}_c\leftarrow\mathrm{MLP}\big(\mathrm{Cat}(\mathbf{e}_{c^1},\dots,\mathbf{e}_{c^L})\big)$ for all $c\in\mathcal{S}$ \Comment{SID-side representations}
  \State Compute token-level score $\sum_{l=1}^{L}\log\bar{p}_{1|r,1}(c^l\,|\,x_r,\mathcal{H},\theta)$ \Comment{Single-step token generation}
  \State Compute SID-level score $\langle \mathbf{u},\mathbf{v}_c\rangle$ for all $c\in\mathcal{S}$
  \State Compute the dual-level flow contrastive loss $\mathcal{L}$ by Eq.~\ref{eq:dual-level-loss}
  \State Update $\theta$ with AdamW
\EndWhile
\State \Return $\theta$
\end{algorithmic}
\end{algorithm}

\vspace{0.5em}

\begin{algorithm}[H]
\caption{Inference of SPRINT.}
\label{alg:inference}
\begin{algorithmic}[1]
\Require User history $\mathcal{H}$; trained parameters $\theta$; the set of valid trajectories in the SID prefix tree $\mathcal{T}$; recommendation list length $K$
\Ensure Ranked list of $K$ items
\State \textbf{Offline:} precompute $\mathbf{v}_c$ for all $c\in\mathcal{T}$ \Comment{User/Sequence-independent, computed once}
\State $x_0\leftarrow[\mathbb{m},\dots,\mathbb{m}]$ \Comment{Generate from all-mask sequence}
\State Run one forward pass of the encoder-decoder on $(\mathcal{H},x_0)$ to obtain $\bar{p}_{1|0,1}(\cdot\,|\,x_0,\mathcal{H},\theta)\in\mathbb{R}^{L\times M}$ and $\mathbf{u}$ \Comment{Single-step generation}
\State Compute the generation score of an SID using Eq.~\ref{eq:inference}
\State \Return the $K$ SIDs with the highest generation probability
\end{algorithmic}
\end{algorithm}
\end{minipage}
\end{figure}

\section{Limitations}
\label{sec:limitations}
This work aims to develop a single-step SID generation paradigm from a new perspective of average probability velocity. It still has several limitations. First, although the average-velocity formulation is defined over a general interval $[r,t]$, SPRINT fixes the terminal time $t=1$ and anchors most training sources at $r=0$. As our experiments show, the best performance is mostly obtained with $\rho\in\{0.75,1\}$, \textit{i.e.}, when the large majority of sources match the strict single-step inference condition, while a small fraction of partially committed sources can still be beneficial. This suggests that intermediate intervals carry useful signal, but we only exploit them as auxiliary supervision for the single map from $r=0$ to $t=1$, and do not explore their behavior at inference or a tunable few-step variant, which we leave to future work. Second, since SPRINT decodes over all valid trajectories in SID prefix tree, its inference cost grows with the size of the item corpus. It is thus best suited to the ranking stage, where a candidate set has already been retrieved and the goal is to produce a fine-grained recommendation list, rather than to fully end-to-end settings that retrieve directly from billions of items. This limitation is also shared by many existing GR works \citep{wang2026one, hu2026empowering, yang2025sparse}. Scaling single-step generation to such settings is left to future work.

\end{document}